\documentclass[journal]{IEEEtran}

\ifCLASSINFOpdf
   \usepackage[pdftex]{graphicx}
   \usepackage{color}
   \usepackage{amsmath,amsfonts}
   \usepackage{bbm}
   \usepackage{amsmath,bm}
   \usepackage{accents}
   \usepackage[noadjust]{cite}

\usepackage{tikz-timing}
\usepackage{tikz}
\usepackage{subfigure}
\usetikzlibrary{arrows,positioning,automata}

   \usepackage{color}

   \usepackage{subfigure}

   \newtheorem{theorem}{Theorem}
   \newtheorem{definition}{Definition}

   \newtheorem{proposition}{Proposition}
   \newtheorem{corollary}{Corollary}
   
   \newtheorem{example}{Example} 

   \newcommand*{\QEDA}{\hfill\ensuremath{\triangle}}
   \usepackage{epstopdf}
   \usepackage{graphicx}
   
   \newcommand{\R}{\mathbb{R}}
   
   \newcommand{\av}{{\rm av}}
   \newcommand{\col}{{\rm col}}

   \newcommand{\RS}{\mathbb{S}}
   \newcommand\SmallMatrix[1]{{%
   		\scriptsize\arraycolsep=0.4\arraycolsep\ensuremath{\begin{bmatrix}#1\end{bmatrix}}}}
   	
   	\newcommand{\dist}{\rm dist}
   	\usepackage{stfloats}

\else

\fi

\begin{document}

\title{Partial Exponential Stability Analysis of Slow-fast Systems via Periodic Averaging}

\author{Yuzhen~Qin,~\IEEEmembership{Student Member,~IEEE}, Yu~Kawano,~\IEEEmembership{Member,~IEEE}, and Brian~D.~O.~Anderson,~\IEEEmembership{Life Fellow,~IEEE}, and Ming~Cao,~\IEEEmembership{Senior Member,~IEEE}\\[2mm]
\thanks{Y. Qin and M. Cao are with Engineering and Technology Institute (ENTEG), University of Groningen, the Netherlands (emails: \{y.z.qin, m.cao\}@rug.nl). Y. Kawano is with the Graduate School of Engineering, Hiroshima University, Higashi-Hiroshima, Japan (email: ykawano@hiroshima-u.ac.jp). B.D.O. Anderson is with School of Automation, Hangzhou Dianzi University, Hangzhou, 310018, China, and Data61-CSIRO and {Research School of Electrical, Energy and Materials Engineering,} Australian National University, Canberra, ACT 2601, Australia (email: Brian.Anderson@anu.edu.au). 	
	The work of M. Cao is supported in part by the European Research Council (ERC-CoG-771687) and the Netherlands Organization for Scientific Research (NWO-vidi-14134). The work of B.D.O. Anderson is supported by the Australian Research Council (ARC) under grants  DP-190100887 and DP-160104500, and by Data61-CSIRO.
}

}


\maketitle

\begin{abstract}
This paper presents some new criteria for partial exponential stability of a slow-fast nonlinear system with a fast scalar variable using periodic averaging methods. Unlike classical averaging techniques, we construct an averaged system by averaging over this fast scalar variable instead of the time variable. We then show that partial exponential stability of the averaged system implies partial exponential stability of the original one. As some intermediate results, we also obtain a new converse Lyapunov theorem and some perturbation theorems for partially exponentially stable systems. We then apply our established  criteria to study remote synchronization of  Kuramoto-Sakaguchi oscillators coupled by a star network with two peripheral nodes. We analytically show that detuning the natural frequency of the central mediating oscillator can increase the robustness of the remote synchronization against phase shifts. 
\end{abstract}


\IEEEpeerreviewmaketitle

\section{Introduction}

\IEEEPARstart{P}{artial} stability describes a property of a dynamical system that only a given part of its states, instead of all, are stable. The earliest study on partial stability dates back to a century ago in Lyapunov's seminal work in 1892, and some comprehensive and well-known results of which were documented by Vorotnikov in his book \cite{vorotnikov1997partial}.
Different from  standard full-state stability theory which usually deals with stability of point-wise equilibria, partial stability is more associated with stability of motions lying in a subspace \cite{haddad2011nonlinear}. It provides a unified and powerful framework to study a range of application-motivated  theoretical problems, such as spacecraft stabilization by rotating masses \cite{vorotnikov1997partial}, inertial navigation systems \cite{sinitsyn1991stability}, transient stability of power systems \cite{willems1974partial}, output regulation \cite{isidori1990output}, and synchronization in complex networks \cite{gambuzza2013analysis,qin2018stability}.

Some Lyapunov criteria have been established to study partial stability of nonlinear systems \cite{vorotnikov1997partial},  \cite[Chap. 4]{haddad2011nonlinear}, \cite{miroshnik2004attractors},   \cite{hancock2014restricted}. However, when it comes to the analysis of multi-time scale systems,  the existing results are usually  difficult to apply. In fact, time scale separation  is pervasive in physical, biological, neuroscientific, and ecological systems \cite[Chap. 20]{kuehn2015multiple}, and one often needs to study the partial stability of them. Therefore, there is a great need to further develop new criteria for partial stability analysis, in particular in the setting of slow-fast systems. 
In this paper, we aim at developing new criteria for exponential partial stability of a particular type of slow-fast systems, wherein the fast variable is a scalar. Various practical systems can be modeled by this type of slow-fast systems,  such as semiconductor lasers \cite{al2009chaotic}, and mixed-mode oscillations in chemical systems \cite{petrov1992mixed}, where the fast scalar variables are the photon density and a chemical concentration, respectively. In particular, fast time-varying systems can always be modeled in this way since the time variable $t$ can be taken as the fast scalar \cite{chellaboina2002unification}. 

In full-state stability analysis of fast time-varying systems, averaging methods are widely used to establish criteria for full-state exponential stability \cite[Chap. 10]{khalil2002nonlinear} \cite{aeyels1999exponential} and also asymptotic stability \cite{tsinias2013averaging}. 
Inspired by these works, we utilize periodic averaging techniques and establish some criteria for partial exponential stability of the considered type of slow-fast systems. 
Unlike what is usually done in standard averaging, we construct an averaged system by averaging the original one over the fast scalar variable, which is in general different from the time variable. We show that partial exponential stability of the averaged system implies partial exponential stability of the original one. Compared to the existing criteria for full-state exponential stability \cite{chellaboina2002unification,khalil2002nonlinear,tsinias2013averaging}, the analysis in our case is much more challenging since some state variables are  unstable. To construct the proof, we also develop a new converse Lyapunov theorem and some perturbation theorems for partially exponentially stable systems. Unlike the converse Lyapunov theorem in \cite[Theorem 4.4]{haddad2011nonlinear}, we present two bounds for the partial derivatives of the Lyapunov function with respect to the stable and unstable states, respectively. Moreover, our obtained perturbation theorems are the first-known ones for partial exponential stability analysis, although their counterparts  \cite[Chap. 9 and 10]{haddad2011nonlinear} for full-state stability have been widely used to analyze perturbed systems. 

We then apply our obtained results to study the partial exponential stability of a concrete slow-fast system that arises from the remote synchronization problem of coupled phase oscillators. Remote synchronization describes the phenomenon that oscillators coupled indirectly get synchronized, but the ones connecting them are not synchronized with them \cite{gambuzza2013analysis}.  
In fact, remote synchronization is ubiquitous in nature. For example, distant cortical regions without apparent direct neural links in the human brain are detected to behave coherently \cite{varela2001brainweb}. Unlike the emergence of the classical synchronization, for which strong connections are required  \cite{qin2019partial,menara2019stability,dorfler2014synchronization}, remote synchronization is more associated with morphological symmetry. For example, nodes located remotely in a network might be able to swap their positions without changing the functioning of the overall system \cite{nicosia2013remote}. In this paper, we consider the perhaps simplest network that is morphologically symmetric (i.e., a star network with two peripheral nodes), and study remote synchronization of oscillators coupled by this network. Despite its simple structure, this fundamentally important  network has been shown to give rise to zero-lag synchronization of remotely separated neuronal populations, even in the presence of considerable synaptic conduction delays \cite{vicente2008dynamical}.  Experiments have also demonstrated that this network can render isochronous synchronization of delay-coupled semiconductor lasers \cite{fischer2006zero}, and chaotic electronic circuits \cite{wagemakers2007isochronous}. The central element in this network, which is dynamically relaying or mediating the dynamics of the peripheral ones, is believed to play an essential role. Some recent works show that the remote synchronization can be enhanced if some parameter mismatch or heterogeneity is introduced to the central element \cite{banerjee2012enhancing,gambuzza2016inhomogeneity}. We seek to analytically study this interesting experimental finding, which is quite challenging.  

Towards this end, we employ the Kuramoto-Sakaguch model to describe the dynamics of the oscillators \cite{sakaguchi1986soluble}. 
Different from the classical  Kuramoto model, there is an additional phase shift term in the Kuramoto-Sakaguchi model. This phase shift is usually used to model small time delays \cite{panaggio2015chimera}, e.g., synaptic connection delays \cite{hoppensteadt2012weakly}. We then detune the natural frequency of the central oscillator to investigate whether the introduction of parameter heterogeneity  can strengthen remote synchronization. When there is no natural frequency detuning, we show that a large phase shift can destabilize remote synchronization.  In sharp contrast, remote synchronization surprisingly becomes more robust against phase shifts when there is sufficiently large natural frequency detuning. Actually, it remains exponentially stable for any phase shift. In other words, we provide a rigorous proof for the first time explaining the  phenomenon experimentally demonstrated by \cite{banerjee2012enhancing,gambuzza2016inhomogeneity}. Compared to the classical synchronization of coupled Kuramoto oscillators where frequency synchronization usually occurs, the analysis is much more challenging since the central oscillator has a different frequency than the peripheral ones when remote synchronization occurs. Modeling the problem into a slow-fast system, we construct the proof using our obtained criteria for partial stability. 

The rest of the paper is structured as follows. Section \ref{sec:2} introduces the model of a slow-fast system and formulates the problem. The main results and corresponding analysis are provided in Section \ref{sec:3}. As an application, remote synchronization of Kuramoto-Sakaguchi oscillators is studied in Section \ref{synchronization:sec}. Brief concluding remarks appear in Section \ref{sec:conclusion}.

\textit{Notations}: Let $\R$ denote the set of real numbers. For any $\delta>0$, let $\mathcal B_\delta:=\{x\in \mathbb R^n:\|x\| < \delta\}$. Given two vectors $x\in \R^n$ and $y\in \R^m$, denote $\col(x,y)=(x^\top,y^\top)^\top$.  
Denote the unit circle by $\RS^1$, and a point on it is called a \textit{phase} since the point can be used to indicate the phase angle of an oscillator. Let $\RS^n= \RS^1\times \cdots \times \RS^1$.

%

\section{Problem Formulation}\label{sec:2}
A wide range of systems exhibit  multi-timescale dynamics, and among them many have a fast changing variable that is scalar. This motivates us to study a class of slow-fast systems in this paper, whose dynamics are described by
\begin{subequations}\label{main}
	\begin{align}
		&\dot x=f_1(x,y,z), \label{main:1}\\
		&\dot y=f_2(x,y,z),\label{main:2}\\
		\varepsilon &\dot z=f_3(x,y,z),\label{main:3}
	\end{align}
\end{subequations}
where $x\in\R^n,y\in\R^m,z\in \R$, and $\varepsilon>0$ is a small constant. That is, $x,y$ are the states of slow dynamics, and $z$ is the state of fast dynamics. All the maps, $f_1:\R^{n+m+1}\to \R^n$, $f_2:\R^{n+m+1} \to \R^m$, $f_3:\R^{n+m+1} \to \R$, are continuously differentiable, and $T$-periodic in $z$, i.e., $f_i(x,y,z+T)=f_i(x,y,z)$ for all $i=1,2,3$. We further assume that the $f_i$'s are such that the solution to the system \eqref{main} exists for all $t\ge 0$. Moreover, $x=0$ is a partial equilibrium point of the system \eqref{main}, i.e., $f_1(0,y,z)=0$ for any $y\in \R^n$ and $z\in \R$. Also, $f_2(0,y,z)=0$ for any $y$ and $z$. However,~$f_3$ is not required to satisfy~$f_3(0,y,z)=0$.

 We are interested in studying partial exponential stability of the system \eqref{main}.  Let us first define uniform partial exponential stability.


\begin{definition}[{\cite[Chap. 4]{haddad2011nonlinear}}] \label{defi:stability}
	A partial equilibrium point $x=0$ of the system \eqref{main} is \emph{exponentially $x$-stable} uniformly in $y$ and $z$ if there exist $c_1,c_2,\delta>0$ such that $\|x({0})\|<\delta$ implies that $ \|x(t)\|\le c_1 { \|x(0)\|} e^ {-c_2t}$ for any $t\ge 0$ and $(y_0,z_0) \in\R^{m} \times \R$. 
\end{definition}

Note that when we refer to this definition, we also say that $x = 0$ of the system \eqref{main} is partially exponentially stable or the system \eqref{main} is partially exponentially stable with respect to $x$, without causing any confusion since only uniform partial exponential stability is studied in this paper. Some noticeable efforts have been made to study partial stability of nonlinear systems,  \cite{vorotnikov1997partial},  \cite[Chap. 4]{haddad2011nonlinear},  \cite{hancock2014restricted}. Although these results do not explicitly utilize the slow-fast structure, it is possible to apply them to slow-fast systems. Some Lyapunov criteria have been established in \cite[Chap. 4]{haddad2011nonlinear}. However, it is not always easy to verify partial stability by using such criteria. As a motivating example, we consider the following academic but suggestive model. 

\begin{example}\label{exam:1}
	Consider a nonlinear system whose dynamics are described by 
	\begin{align*}
		&\dot x=-x-0.2x\sin y-2x\cos z,\\
		&\dot y=2x\cos y+x\sin z,\\
		\varepsilon&\dot z=3-\sin x+\cos y.
	\end{align*}
	As it will be shown later, for sufficiently small $\varepsilon>0$, it is  possible to prove that the partial equilibrium point $x=0$ is exponentially stable uniformly in $y$ and $z$. However, it is difficult to construct an appropriate Lyapunov function using the criteria in \cite[Chap. 4]{haddad2011nonlinear}. For example, we might reasonably  choose $V=x^2$ as a Lyapunov function candidate. Its time derivative is $\dot V=-2(1+0.2\sin y+2\cos z)x^2$, which can be positive for some $y$ and $z$, while it is required by \cite[Theorem 1, Chap. 4]{haddad2011nonlinear} to be negative for any $x\neq0$, $y$, and $z$ in order to show the partial exponential stability.   
	{\hfill \QEDA}		
\end{example}

Motivated by the above example, in the next section we aim at further developing Lyapunov theory for partial stability analysis of slow-fast systems.

\section{Partial Stability of Slow-Fast Dynamics}\label{sec:3}

In this section, our goal is to provide a new Lyapunov criterion for partial stability of slow-fast systems. Our analysis consists of several steps. We first construct reduced slow dynamics. Under some practically reasonable assumptions, the partial stability of the constructed slow dynamics and the original slow-fast system are shown to be equivalent. That is, analysis reduces to the partial stability analysis of the constructed slow dynamics. Moreover, since the original slow-fast system is periodic, the constructed one is also periodic.

Next, in order to study the partial stability of the constructed slow dynamics, we use an averaging method. For periodic systems, averaging methods have been widely used to establish criteria for the standard full-state exponential stability \cite[Chap. 10]{khalil2002nonlinear} \cite{aeyels1999exponential} and also asymptotic stability \cite{tsinias2013averaging}. Inspired by these works, we will develop a new criterion for partial stability of the fast periodic dynamics via averaging. According to our new criterion, if the averaged system is partially exponentially stable, then the slow dynamics and consequently the original periodic slow-fast system is partially exponentially stable for sufficiently small $\varepsilon>0$. It is worth emphasizing that compared with the case of full-state stability, partial stability analysis is much more challenging since some states are not stable.

\subsection{Slow Dynamics}
As the first step, we construct a reduced slow dynamics studied in the following subsections. One important fact is that the partial stability of the constructed slow dynamics is equivalent to that of the original slow-fast system~\eqref{main} under the assumption below.

Assume that for the fast subsystem:
\begin{align}
	&f_3(x,y,z)\ge \vartheta, &\forall x\in\R^n,y\in\R^m,z\in \R, \label{f3:bound}
\end{align}
or $f_3(x,y,z)\le -\vartheta$, where~$\vartheta >0$. Note that we only consider the first inequality in this paper since these two inequalities are essentially the same. This assumption \eqref{f3:bound} is naturally satisfied for some practical problems such as vibration suppression of rotating machinery where $f_3$ is the angular velocity \cite{adams2009rotating}, and spin stabilization of spacecrafts where $f_3$ describes the spin rate \cite{wertz2012spacecraft}.

The assumption~\eqref{f3:bound} implies that~$t \mapsto z(t)$ can be interpreted as a change of time (recall that $z$ is a scalar). In fact, \eqref{f3:bound} implies that for any given initial state $(x(0),y(0),z(0))$ of the slow-fast system~\eqref{main}, a part of the solution $z(t)$ is a strictly increasing function of $t$. That is,~$t \mapsto z(t)$ is a global diffeomorphism{\footnote{This follows from the earlier assumption that the solution to the system \eqref{main} exists for all $t\ge 0$.}} from $[0, \infty)$ to $[0, \infty)$. In the new time axis~$z(t)$, the slow-fast system becomes
\begin{align*}
	&\frac{d x(t)}{dz(t)} = \frac{d x(t)}{dt} \frac{dt}{dz(t)} = \varepsilon \frac{f_1 (x(t),y(t),z(t))}{f_3(x(t),y(t),z(t))},\\
	&\frac{d y(t)}{dz(t)} = \frac{d y(t)}{dt} \frac{dt}{dz(t)} = \varepsilon \frac{f_2 (x(t),y(t),z(t))}{f_3(x(t),y(t),z(t))},\\
	&\frac{d z(t)}{dz(t)} = 1,
\end{align*}
The first two subsystems can be viewed as time-varying systems with the new time variable~$z(t)$. Note that, since~$t \mapsto z(t)$ is a global diffeomorphism, the partial stability with respect to~$x$ of the first two time-varying subsystems in the new time axis is equivalent to the partial stability with respect to~$x$ of the system~\eqref{main} in the original time axis. Therefore, hereafter we focus on the first two time-varying subsystems in the new time axis. For the sake of simplicity of description, the first two time-varying subsystems are rewritten into
\begin{subequations}\label{reduced}
	\begin{align}
		&\frac{dx}{dz}=\varepsilon h_1(x,y,z),\label{reduced:1}\\
		&\frac{dy}{dz}=\varepsilon h_2(x,y,z),\label{reduced:2}
	\end{align}
\end{subequations}
where $h_1=f_1/f_3$, and $h_2=f_2/f_3$. From the properties of $f_1,f_2$, and $f_3$, it follows that both $h_1(0,y,z)=0$ and $h_2(0,y,z)=0$ for any $y\in \R^m$ and $z\in \R$, and the solution to the system \eqref{reduced} exists for all $z \ge 0$.  Moreover, the constructed slow dynamics~\eqref{reduced} is again~$T$-periodic in~$z$.

\subsection{Partial Stability Conditions via Averaging}

In order to study partial stability with respect to~$x$ of the constructed periodic slow dynamics~\eqref{reduced}, we use an averaging technique for periodic systems. The averaged system obtained from the slow dynamics~\eqref{reduced} can be used for the partial stability analysis of the slow dynamics~\eqref{reduced} before averaging. From the discussion in the previous subsection, the partial stability of the slow dynamics~\eqref{reduced} is equivalent to that of the original slow-fast system~\eqref{main}.

Since the slow dynamics~\eqref{reduced} is~$T$-periodic in~$z$, it is possible to apply an averaging method. We associate it with a  partially averaged system given by
\begin{subequations}\label{average}
	\begin{align}
	&\frac{d w}{dz}=\varepsilon h_{\av}(w,v),\label{average:1}\\
	&\frac{dv}{dz}=\varepsilon h_2(w,v,z),\label{average:2}
	\end{align}
\end{subequations}    
where the function $h_{\av}$ is defined by
\begin{align}\label{expre:hav} 
h_\av (w,v) = \frac{1}{T} \int_{0}^{T}h_1(w,v,\tau)d\tau,
\end{align} 
 where~$h_\av (0,v)=0$ for any~$v \in \R^m$ from~$h_1(0,v,z)=0$.
Note that only the dynamics of~$w$ is averaged with respect to~$z$. 

In fact, if the averaged system~\eqref{average} is partially exponentially stable with respect to~$w$, then the periodic slow system~\eqref{reduced} is partially exponentially stable with respect to~$x$ for sufficiently small $\varepsilon>0$.	This implies that partial exponential stability of the original slow-fast system~\eqref{main} can be verified by using the averaged system~\eqref{average}. This fact is stated formally as follows, which is one of the main results in this paper.

\begin{theorem}\label{theo:4}
	Suppose that $w=0$ of the averaged system~\eqref{average} is partially exponentially stable uniformly in $v$, i.e., there exists $\delta>0$ such that for any~$z_0 \in \R$ and $w(0)\in\mathcal{B}_\delta$,
	\begin{align}
	&\|w(z)\|\le k\|w(0)\|e^{-\lambda (z-z_0)},& k,\lambda>0, \forall z\ge z_0. \label{aver:expon}
	\end{align}
	Assume that there are $L_1,L_2> 0$ such that for any $x\in \mathcal B_\delta, y\in \R^m,z\in \R$, the functions $h_1$ and $h_2$ in \eqref{reduced} satisfy
	\begin{align}
	&\left\|\frac{\partial h_1}{\partial x}(x,y,z)\right\|\le L_1,&\left\|\frac{\partial h_2}{\partial x}(x,y,z)\right\|\le L_2.\label{aver:bound}
	\end{align}

	Then, there exists $\varepsilon_1>0$ such that, for any $\varepsilon<\varepsilon_1$, the partial equilibrium point $x=0$ of the system \eqref{reduced} is exponentially stable uniformly in $y$. As a consequence, for any $\varepsilon<\varepsilon_1$, the partial equilibrium point $x=0$ of the system \eqref{main} is exponentially stable uniformly in $y$ and $z$. 	{\hfill \QEDA}
\end{theorem}

The following subsections are dedicated to proving this theorem. Before providing the proof, we illustrate its utility. First, let us look back at Example~\ref{exam:1}, and see how the obtained results can be applied.

\textit{Continuation of Example \ref{exam:1}}:
As $3-\sin x+\cos y\ge 1$ for any $x,y$, the property~\eqref{f3:bound} holds. Then, one can construct the averaged system~\eqref{average} of the system in Example \ref{exam:1} as
	\begin{align*}
	&\frac{dw}{dz}=\varepsilon\frac{-w-0.2w\sin v}{3-\sin w+\cos v},\\
	&\frac{dv}{dz}=\varepsilon\frac{2w\cos v+w\sin z}{3-\sin w+\cos v}.
	\end{align*}
Choose a Lyapunov function candidate $V(w,v,z)=w^2$. Then, it holds that 
	\begin{align*}
	\frac{dV}{dz} = -2 \varepsilon \cdot \frac{1+0.2\sin v}{3-\sin w+\cos v}w^2\le -\frac{8}{15}\varepsilon w^2.
	\end{align*}
According to \cite[Theorem 1]{haddad2011nonlinear}, $w=0$ of the averaged system is partially exponentially stable. From Theorem \ref{theo:4}, one can conclude that  $x=0$ of the original system  in Example \ref{exam:1} is partially exponentially stable if $\varepsilon>0$ is  sufficiently small.	{\hfill \QEDA}
\\

By using averaging techniques, Theorem \ref{theo:4} provides a new way to study partially stability of slow-fast systems for which the existing criteria are difficult to apply. As  another application, we can recover the conventional criteria \cite[Chap. 10]{khalil2002nonlinear} and \cite{aeyels1999exponential} for exponential stability of fast time-varying systems, as follows. Consider  the following system with respect to~$x$,	
	\begin{subequations}\label{simplified}
		\begin{align}
		&\dot x=f_1(x,z), \\
		&\varepsilon \dot z=f_3(x,z),
		\end{align}
	\end{subequations}
where $f_1$ and $f_3$ satisfy all the assumption made for system~\eqref{main}. The difference from~\eqref{main} is that there is no variable~$y$. To study partial stability with respect to~$x$, we apply the change of time-axis, $t \to z$. Then, we have
	\begin{align}
		\frac{dx}{dz}=\varepsilon h_1(x,z) \label{simplified_reduced:1},
	\end{align}
	where $h_1:=f_1/f_3$.
Next, compute the averaged system of the fast subsystem
\begin{align}\label{simplified:aver}
&\frac{dw}{dz} = \varepsilon {\hat h}_{\av}(w),
\end{align} 
where the function ${\hat h}_{\av}$ is defined by
\begin{align}\label{simplified:hav}
{\hat h}_\av (w) = \frac{1}{T} \int_{0}^{T} {h}_1(w,\tau)d\tau,
\end{align} 
As expected, if the averaged system~\eqref{simplified:aver}  is exponentially stable, then the partial stability of \eqref{simplified} is ensured as long as $\varepsilon>0$ is sufficiently small, which is formally stated in the following corollary. If $f_3(x,z)=1$ for all $x\in \R^n$ and $z \in \R$, this corollary reduces to the criteria in \cite[Chap. 10]{khalil2002nonlinear} and \cite{aeyels1999exponential} for exponential stability of fast time-varying systems since $z$ and $t$ are the same.
\begin{corollary}\label{Coro:2:aver}
	Suppose that $w=0$ is exponentially stable for the averaged system \eqref{simplified:aver}.
	Assume that there is $L> 0$ such that for any $x\in \mathcal B_\delta, z\in \R$ the function $h_1$ in \eqref{simplified_reduced:1} satisfies
	\begin{align}
	&\left\|\frac{\partial h_1}{\partial x}(x,z)\right\|\le L.\label{simplified:aver:bound}
	\end{align}
	Then, there exists $\varepsilon_1>0$ such that, for any $\varepsilon<\varepsilon_1$, the partial equilibrium $x=0$ of the system \eqref{simplified} is partially exponentially stable uniformly in $z$. 	{\hfill \QEDA}\end{corollary}

In next section, we show how our results on partial exponential stability can be applied to remote synchronization in a simple network of Kuramoto oscillators. Before that, we construct the proof of Theorem~\ref{theo:4} in the following subsections.

\subsection{Preparatory Results for Proving Theorem~\ref{theo:4}}
In the following subsections, our objective is to prove Theorem~\ref{theo:4}, that is to show the partial exponential stability of the averaged system~\eqref{average} implies that of the periodic slow system~\eqref{reduced}. Recall that partial stability of the slow system~\eqref{reduced} is equivalent to that of the original slow-fast system~\eqref{main} under~\eqref{f3:bound}.

For conventional full-state exponential stability analysis, the original system is regarded as a perturbed system of the averaged one. As long as the perturbation characterized by~$\varepsilon$ is sufficiently small, the exponential stability of the original system is ensured \cite{aeyels1999exponential}, \cite[Chap. 10]{khalil2002nonlinear}. Similar ideas are used in this paper for partial exponential stability analysis. Instead of full-state stability, we only require the averaged system \eqref{average} to be partially exponentially stable.

In order to show the partial exponential stability of the periodic slow system~\eqref{reduced}, we use Lyapunov theory. First, we construct a Lyapunov function for a partially exponentially stable averaged system. Then, by using this Lyapunov function, we show the partial exponential stability of the periodic slow system if the perturbation is sufficiently small. This subsection is dedicated to constructing a Lyapunov function. In effect, we provide a new converse Lyapunov theorem for partial stability.

As a generalized form of \eqref{average},  we consider the following time-varying systems in this section
\begin{subequations}\label{side:main}
	\begin{align}
	\frac{dw}{dz}&=\varphi_1(w,v,z), \label{side:1}\\
	\frac{dv}{dz}&=\varphi_2(w,v,z), \label{side:2}
	\end{align}
\end{subequations}
where~$w\in\R^n$,~$v\in\R^m$,~$z \in \R$, and the functions, $\varphi_1:\R^{n+m+1} \to \R^n$, $\varphi_2:\R^{n+m+1}\to \R^m$ are continuously differentiable. Moreover, it holds that $\varphi_1(0,v,z)=0$ and $\varphi_2(0,v,z)=0$ for any $v\in \R^m$. We further assume that for any $z_0$ the solution to the system \eqref{side:main} exists for all $z\ge z_0$.

Now, we provide a converse theorem for exponential partial stability of the system~\eqref{side:main}, which is directly applicable to the averaged system~\eqref{average}. 
\begin{theorem}\label{Theo:converse}
Suppose that $w=0$ is partially exponentially stable uniformly in $v$ for the system \eqref{side:main}, i.e., there exists $\delta>0$ such that for any~$z_0 \in \R$ and $w(0)\in\mathcal{B}_\delta$, 
	\begin{align}
	&\|w(z)\|\le k\|w(0)\|e^{-\lambda (z-z_0)},&k,\lambda>0, \forall z\ge z_0. \label{expo:decay}
	\end{align}
Also, assume that there are $L_1,L_2> 0$ such that
	\begin{align}
	&\left\|\frac{\partial \varphi_1}{\partial w}(w,v,z)\right\|\le L_1, &\left\|\frac{\partial \varphi_2}{\partial w}(w,v,z)\right\|\le L_2,\label{fistpart:bound}
	\end{align}	
	for any $w\in \mathcal B_\delta, v\in \R^m, z\in \R$.
	Then, there exists a function $V:  \mathcal B_\delta \times \R^m \times \R\to \R$ that satisfies the following inequalities:
	\begin{align}
	&c_1\|w\|^2 \le V(w,v,z)\le c_2\|w\|^2, \label{conver:bounds}\\
	&\frac{\partial V}{\partial z} + \frac{\partial V}{\partial w}\varphi_1(w,v,z)+\frac{\partial V}{\partial v}\varphi_2(w,v,z)  \le -c_3\|w\|^2, \label{conver:decay}\\
	&\left\| \frac{\partial V}{\partial w}\right\| \le c_4\|w\|,\label{parti-V:bounds}\\
	&	\left\| \frac{\partial V}{\partial v}\right\| \le c_5\|w\| \label{parti-V:bounds:y},
	\end{align}
	for some positive constants $c_1,c_2,c_3,c_4$ and $c_5$. {\hfill \QEDA}
\end{theorem}

The same uniform boundedness assumptions on the partial derivatives of the functions $\varphi_1$ and $\varphi_2$ are actually made in Theorem \ref{Theo:converse} and Theorem 4.4 of \cite{haddad2011nonlinear}. Unlike the theorem in \cite{haddad2011nonlinear}, we work on time-varying systems. Moreover, we obtain additional two bounds for the partial derivative of $V$, namely~\eqref{parti-V:bounds} and \eqref{parti-V:bounds:y} by assuming $\varphi_2(0,v,z)=0$ for any $v$ and $z$. Thus, our proof is more involved; it can be found in Appendix \ref{proof_Theo:2}.
	
\subsection{Analysis of Perturbed Systems}

In the previous subsection, we have constructed a converse Lyapunov theorem. By applying this to a partially exponentially stable averaged system~\eqref{average}, one can construct a Lyapunov function satisfying all conditions in Theorem~\ref{Theo:converse}. By using this Lyapunov function, we consider to study the partial exponential stability of the periodic slow dynamics~\eqref{reduced}. As typically done in averaging methods, we consider  periodic  slow dynamics~\eqref{reduced} as a perturbed system of its averaged system~\eqref{average}. Then, we conclude the partial exponential stability of the periodic slow dynamics~\eqref{reduced} by using the Lyapunov function for its averaged system~\eqref{average}. 

To this end,  we study the following perturbed system of the system~\eqref{side:main} in the previous subsection:
 \begin{subequations}\label{side:perturbed}
	\begin{align}
	\frac{d w_p}{dz} &=\varphi_1(w_p,v_p,z)+g_1(w_p,v_p,z), \label{pertirbed:1}\\
	\frac{d v_p}{dz} &=\varphi_2(w_p,v_p,z)+g_2(w_p,v_p,z), \label{pertirbed:2}
	\end{align}
\end{subequations} 
where $g_1:\R^{n+m+1} \to \R^{n}$ and $g_2:\R^{n+m+1}\to \R^{m}$ are piecewise continuous in $z$ and locally Lipschitz in $(w_p,v_p)$.  Particularly, we assume that the perturbation terms satisfy the bounds
\begin{align}
&\|g_1(w_p,v_p,z)\|\le \gamma_1 (z)\|w_p\|+\psi_1(z), \label{pert:1}\\
&\|g_2(w_p,v_p,z)\|\le \gamma_2 (z)\|w_p\|+\psi_2(z),\label{pert:2}
\end{align}
where $\gamma_1,\gamma_2:\R \to \R$ are nonnegative and continuous for all $z\in \R$, and $\psi_1,\psi_2:\R\to \R$ are nonnegative, continuous and bounded for all $z\in \R$. Notice that the bounds on the right are independent of $v_p$.

The following theorem presents some results on the asymptotic behavior of the perturbed system \eqref{side:perturbed} when the nominal system \eqref{side:main} has a partially exponentially stable equilibrium $w_p=0$. The proof is based on the constructed Lyapunov function in Theorem \ref{Theo:converse}; for the proof, see Appendix~\ref{proof:the03}.

\begin{theorem}\label{Theo:asymBeha}
Suppose that the nominal system \eqref{side:main} satisfies all the assumptions in Theorem \ref{Theo:converse}. Also, assume that the perturbation terms $g_1(w_p,v_p,z)$ and $g_2(w_p,v_p,z)$ are respectively bounded as in \eqref{pert:1} and \eqref{pert:2} for $\gamma_1,\gamma_2$ and $\psi_1,\psi_2$ satisfying the following inequalities
	\begin{align} \label{Theo3:Asp1}
	c_4\int_{z_0}^{z}\gamma_1(\tau )d\tau+c_5 \int_{z_0}^{z}\gamma_2(\tau )d\tau\le \kappa (z-z_0)+\eta,
	\end{align}
	where 
	\begin{align}\label{Theo3:Asp2}
	&0 \le \kappa < \frac{c_1c_3}{c_2}, &\eta \ge 0;
	\end{align}
	and
	\begin{align}\label{Theo3:Asp3}
	& c_4 \psi_1(z)+c_5\psi_2(z) < \frac{2c_1k_1 \delta}{k_2}, &\forall z \ge z_0,
	\end{align}
	where 
	\begin{align}\label{expre:k_1,k_2}
	&k_1=\frac{c_3}{2c_2}-\frac{\kappa}{2c_1}, &k_2=\exp \left( \frac{\eta}{2c_1} \right).
	\end{align}
	Then, the solution to the perturbed \eqref{side:perturbed} satisfies
	\begin{align}\label{asympto_behav}
	\|w_p(z)\| \le &k_2 \sqrt{\frac{c_2}{c_1}}\|w_p(z_0)\|e^{-k_1(z-z_0) } \nonumber\\
	&+\frac{k_2}{2c_1}\int_{z_0}^{z}e^{-k_1(z-\tau) }\psi(\tau)d\tau, &\forall z \ge z_0.
	\end{align}
	for any initial time~$z_0 \in \R$ and any initial state $w_p(z_0)\in \R^{n}$ and $v_p(z_0) \in \R^{m}$ such that
	\begin{align}\label{init:x0}
	\|w_p(z_0)\|<  \frac{\delta}{k_2}\sqrt{\frac{c_1}{c_2}}.
	\end{align}
\end{theorem}
 {\hfill \QEDA}	

A similar result is found in \cite[Lemma 9.4]{khalil2002nonlinear}, where the nominal system is assumed to be exponentially stable. With some perturbation, the asymptotic behavior of the full state is reported there. In contrast, the nominal system studied here is only assumed to be partially exponentially stable in Theorem \ref{Theo:asymBeha}, and we show that the asymptotic behavior of a part of the states $w_p$ for the perturbed system follows some specific rule, without being  concerned about how the other part of the states, $v_p$, is evolving.
	We next consider a particular case where $g_1$ and $g_2$ in \eqref{side:perturbed} are vanishing perturbations, i.e., $\psi_1(z)$ and $\psi_2(z)$ in \eqref{pert:1} and \eqref{pert:2} satisfy $\psi_1(z)=\psi_2(z)=0$, and obtain the next corollary. In fact, this corollary is used to prove Theorem~\ref{theo:4}.

\begin{corollary}\label{corollary:expenSta}
Suppose that the nominal system \eqref{side:main} satisfies all the assumptions in Theorem \ref{Theo:asymBeha}.  Also, assume that the perturbation terms $g_1(w_p,v_p,z)$ and $g_2(w_p,v_p,z)$ are respectively bounded by $\|g_1(w_p,v_p,z)\|\le \gamma_1 (z)\|w_p\|$ and $\|g_2(w_p,v_p,z)\|\le \gamma_2 (z)\|w_p\|$ for $\gamma_1$ and $\gamma_2$ satisfying \eqref{Theo3:Asp1} and \eqref{Theo3:Asp2}, i.e. $\psi_1(\cdot)=0$ and $\psi_2(\cdot)=0$. Then, $w_p=0$ of the system \eqref{side:perturbed} is partially exponentially stable uniformly in $v_p$. Moreover, the solution to \eqref{side:perturbed} satisfies the following specialisation of \eqref{asympto_behav}:
	\begin{align*}
	&\|w_p(z)\| \le k_2 \sqrt{\frac{c_2}{c_1}}\|w_p(z_0)\|e^{-k_1(z-z_0)}, &\forall z \ge z_0,
	\end{align*}
for any initial time~$z_0 \in\R$ and any initial condition $w_p(z_0)\in \R^{n}$ and~$v_p(z_0) \in \R^{m}$ satisfying \eqref{init:x0}.
\end{corollary}

In the next subsection, we show how the results obtained in the previous and this subsection enable us to use  averaging techniques to study the partial stability of a periodic slow dynamics~\eqref{reduced} from its averaged system~\eqref{average}. 

\subsection{Proof of Theorem~\ref{theo:4}}
Now, we are ready to provide the proof of Theorem~\ref{theo:4}.
\begin{IEEEproof}[Proof of Theorem~\ref{theo:4}]
First, in order to describe the original slow system \eqref{reduced} as a perturbation of the averaged system~\eqref{average}, we introduce the following change of variables  into the original slow system \eqref{reduced}:
\begin{subequations}\label{variable:change}
	\begin{align}
	&x= w_p+\varepsilon u (w_p,v_p,z),\label{variabCha:1}\\
	&y= v_p,\label{variabCha:2}
	\end{align}
\end{subequations}
where 
\begin{align}\label{defi:u}
u(w_p,v_p,z)=\int_{0}^{z} \Delta(w_p,v_p,\tau)d\tau,
\end{align}
with $\Delta(w_p,v_p,z)=h_1(w_p,v_p,z)-h_\av(w_p,v_p)$. From the definition of $h_\av$ in \eqref{expre:hav}, it holds that
\begin{align}\label{Delta:0mean}
\int_{0}^{T}\Delta(w_p,v_p,\tau)d\tau=0.
\end{align}
After substituting~\eqref{variable:change} into \eqref{reduced}, we obtain the following
	\begin{align*}
	&\frac{dx}{dz} = \frac{dw_p}{dz}+\varepsilon \frac{\partial u}{\partial z} +\varepsilon \frac{\partial u}{\partial w_p}\frac{dw_p}{dz} + \varepsilon \frac{\partial u}{\partial v_p} \frac{dv_p}{dz},\\
	&\frac{dy}{dz}=\frac{dv_p}{dz}.
	\end{align*}
	Substituting \eqref{reduced} and \eqref{variabCha:1} into the above equations yields
	\begin{align} \label{derav:changed}
	&P(\varepsilon)\left[ {\begin{array}{*{20}{c}}
		{\frac{dw_p}{dz}}\\
		{\frac{dv_p}{dz}}
		\end{array}} \right] = \nonumber \\
	  &\left[ {\begin{array}{*{20}{c}}
		{\varepsilon {h_1}(w_p+\varepsilon u,v_p,{ z}) - \varepsilon {h_1}\left( w_p,v_p,{ z} \right) + \varepsilon {h_{av}}\left( w_p,v_p \right)}\\
		{\varepsilon {h_2}(w_p+\varepsilon u,v_p,{ z})}
		\end{array}} \right],
	\end{align}
	where 
	\begin{align*}
	P(\varepsilon)=\left[ {\begin{array}{*{20}{c}}
		{I + \varepsilon \frac{{\partial u}}{{\partial w_p}}}&{\varepsilon \frac{{\partial u}}{{\partial y}}}\\
		0&I
		\end{array}} \right].
	\end{align*}
We then show that the obtained dynamics~\eqref{derav:changed} can be viewed as a perturbation of the averaged system~\eqref{average}. Therefore, Corollary~\ref{corollary:expenSta} can be used in order to show the partial stability of the obtained dynamics from that of the averaged system. Our goal is to show that the partial stability of the obtained dynamics~\eqref{derav:changed} implies that of the original slow system \eqref{reduced}.

Let us represent the obtained dynamics~\eqref{derav:changed} by a perturbation of the averaged system~\eqref{average}. For $k=1,2$, let $h_{k}^ i$ be the $i$th component of $h_k$. From the mean value theorem, for each $k=1,2$, there exists $\lambda_{k}^i = \lambda_{k}^i ( w_p,v_p, z, \varepsilon)>0$ such that 
	\begin{align*}
	 &{h_{k}^i}(w_p+\varepsilon u,v_p,{ z}) -{h_{k}^i}\left( {w_p,v_p,{ z}} \right)\\
	 & =\frac{\partial h_k^i}{\partial w_p}(w_p+\varepsilon \lambda_k^i u ,v_p,{ z})\cdot \varepsilon u.
	\end{align*}
	Let  us denote
	\begin{align*}
   &H_1(w_p,v_p,{ z},\varepsilon u) \\
   &=
\left[\frac{\partial h_1^ 1}{\partial w_p}(w_p+\varepsilon \lambda_1^ 1 u ,v_p,{ z}),\dots,\frac{\partial h_1^n}{\partial w_p}(w_p+\varepsilon \lambda_1^n u ,v_p,{ z})\right]^\top\\
  & H_2(w_p,v_p,{ z},\varepsilon u) \\
  &=
\left[\frac{\partial h_2^1}{\partial w_p}(w_p+\varepsilon \lambda_2^1 u ,v_p,{ z}),\dots,\frac{\partial h_2^n}{\partial w_p}(w_p+\varepsilon \lambda_2^n u ,v_p,{z})\right]^\top.
   \end{align*}
   Then we have 
	\begin{align}
	&{h_{1}}(w_p+\varepsilon u,v_p,{z}) -{h_1}\left( {w_p,v_p,{ z}} \right) \nonumber\\
	&= H_1(w_p,v_p,{ z},\varepsilon u) \cdot \varepsilon u, \label{meanval:1}\\
	&{h_{2}}(w_p+\varepsilon u,v_p,{ z}) -{h_2}\left( {w_p,v_p,{ z}} \right) \nonumber\\
	&= H_2(w_p,v_p,{z},\varepsilon u) \cdot \varepsilon u, \label{meanval:2}
	\end{align}
	where both $H_1(w_p,v_p,{z},\varepsilon u)$ and $H_1(w_p,v_p,{z},\varepsilon u)$ are bounded since from \eqref{aver:bound} each $\partial h_k^i/\partial w$ is. 	
	Due to the boundedness of $\|\partial u/\partial z\|$, $\|\partial u/\partial w_p\|$, and $\|\partial u/\partial v_p\|$ from Proposition \ref{u:buound} in Appendix~\ref{proposition:02}, it is clear that the matrix $P(\varepsilon)$ is nonsingular for sufficiently small $\varepsilon>0$, and its inverse can be described as $P^{-1}(\varepsilon)=I+{\mathcal O}(\varepsilon)$ with some ${\mathcal O}(\varepsilon)$. Applying this fact together with the equalities \eqref{meanval:1} and \eqref{meanval:2} to \eqref{derav:changed},  one can show that there are bounded $H'_1(w_p,v_p,{ z},\varepsilon u)$ and $H'_2(w_p,v_p,{ z},\varepsilon u)$ such that
	\begin{subequations}\label{average:perturbed}
	\begin{align}
	\frac{dw_p}{dz} &= \varepsilon {h_{av}}\left( w_p,v_p \right)+\varepsilon ^2 H'_1(w_p,v_p,{z},\varepsilon u)u,\\
	\frac{dv_p}{dz} &=\varepsilon h_2(w_p,v_p,{z})+\varepsilon ^2 H'_2(w_p,v_p,{z},\varepsilon u)u.
 	\end{align}
 	\end{subequations}
This is a perturbation of the averaged system~\eqref{average}.
	
Next, we apply Corollary~\ref{corollary:expenSta} to show that the partial exponential stability of the averaged system~\eqref{average} implies that of its perturbation~\eqref{average:perturbed} for sufficiently small~$\varepsilon>0$. From the definition of $h_\av$, we have
 	\begin{align}\label{par_h_av}
 	\left\|\frac{\partial h_\av}{\partial w_p}(w_p,v_p) \right\|=\left\|\frac{1}{T} \int_{0}^{T}\frac{\partial h_1}{\partial w_p}(w_p,v_p,\tau)d\tau\right\|\le L_1  
 	\end{align}
 	for any $w_p\in \mathcal B_\delta, v\in \R^m$. By the assumption \eqref{aver:bound}, $\partial h_2/\partial w_p\le L_2$. Therefore, both inequalities in \eqref{fistpart:bound} are satisfied. Since the system \eqref{average} is assumed to be partially exponentially stable, all the assumptions in Theorem \ref{Theo:asymBeha} are satisfied. To apply Corollary~\ref{corollary:expenSta}, it remains to show that  the perturbation terms are bounded linearly in $\|w_p\|$. Let $b_1>0$ and $b_2>0$ be constants such that $\|H'_1(w,v,{z},\varepsilon u)\| \le b_1$ and $\|H'_2(w_p,v_p,{z},\varepsilon u)\| \le b_2$. From \eqref{bound:u} in Appendix \ref{proposition:02}, it holds that $\|u(w_p,v_p,s)\|\le 2TL_1 \|w\|$, and then the perturbation terms satisfy 
 	\begin{align*}
 	&\|\varepsilon^2 H'_1(w_p,v_p,z,\varepsilon u)u\| \le 2\varepsilon^2 b_1 TL_1 \|w_p\|,\\
 	&\|\varepsilon^2 H'_2(w_p,v_p,z,\varepsilon u)u\| \le 2\varepsilon^2 b_2 TL_1 \|w_p\|.
 	\end{align*}
Moreover, for sufficiently small $\varepsilon_1>0$, any $\varepsilon<\varepsilon_1$ makes sure both inequalities~\eqref{Theo3:Asp1} and \eqref{Theo3:Asp2} are satisfied. Therefore, Corollary \ref{corollary:expenSta} implies that $w_p=0$ is partially exponentially stable for  the perturbation~\eqref{average:perturbed}, or equivalently, ~\eqref{derav:changed}. In other words, there are $\delta'>0$ and $k',\lambda'>0$ such that $w_p(z_0)\in \mathcal B_{\delta'}$ implies $\|w_p(z)\|\le k'\|w_p(z_0)\|e^{-\lambda' (z-z_0)}$, for all~$z\ge z_0$. 

Finally, we show that the partial exponential stability of the system~\eqref{derav:changed} implies that of the  slow dynamics~\eqref{reduced}. From \eqref{variabCha:1} and \eqref{bound:u} in the Appendix, one obtains
 	\begin{align*}
 	|1-2\varepsilon TL_1|\cdot\|w_p(z)\|\le \|x(z)\| \le |1+2\varepsilon TL_1|\cdot \|w_p(z)\|, 
 	\end{align*}
 	for all $z\ge z_0$. Then, it follows that
 	\begin{align*}
 	&\|x(z)\|\le k' \frac{|1+2 \varepsilon TL_1|}{|1-2 \varepsilon TL_1|}\|x(z_0)\| e^{-\lambda' (z-z_0)}, &\forall z \ge z_0,
 	\end{align*}
 	proving the partial exponential stability of $x=0$ for the system \eqref{reduced} for sufficiently small~$\varepsilon>0$. Finally, one can conclude that $x=0$ is also partially exponentially stable for the original slow-fast system \eqref{main} uniformly in $y$ and $z$ under asssumption~\eqref{f3:bound}.
\end{IEEEproof}

\section{Remote Synchronization in a Network of Kuramoto Oscillators}\label{synchronization:sec}

In this section, we apply the results on partial stability analysis to studying remote synchronization of oscillators.  Remote synchronization is a phenomenon arising in coupled networks of oscillators when two oscillators without a direct connection become synchronized without requiring the intermediate ones on a path linking the two oscillators to also be synchronized with them \cite{gambuzza2013analysis}.  We restrict our attention to remote synchronization in a network motif shown in Fig. \ref{Fig:3Node}. This network is simple, but has been shown to surprisingly account for the emergence of zero-lag synchronization in remote cortical regions of the brain, even in the presence of large synaptic conduction delays \cite{vicente2008dynamical}. Experiments have also evidenced that the same network can give rise to isochronous synchronization of delay-coupled semiconductor lasers \cite{fischer2006zero}. The central element $0$ in this network plays a critical role in mediating or relaying the dynamics of the peripheral $1$ and $2$. A recent work reveals that detuning the parameters of the central element from those of the peripheral ones can actually enhance remote synchronization \cite{banerjee2012enhancing}. To study this interesting finding analytically, we employ Kuramoto-Sakaguchi model  \cite{sakaguchi1986soluble} and detune the natural frequency of the central oscillator.

\subsection{Problem Statements}
Using Kuramoto-Sakaguchi model and detuning the natural frequency of the central oscillator, the dynamics of the oscillators are described by
\begin{subequations}\label{dyna:3osci}
	\begin{align}
		&\dot \theta_i=\omega + A_i\sin(\theta_0-\theta_i-\alpha),i ={1,2}; \label{dyna_peripheral}\\
		&\dot \theta_0 = \omega +\sum_{j=1}^{2}A_j\sin (\theta_j-\theta_0-\alpha) + u, \label{dyna_central}
	\end{align}
\end{subequations}
where $\theta_i\in \mathbb{S}^1$ is the phase of the $i$th oscillator; $\omega>0$ is the uniform natural frequency of each oscillator; $A_i>0$ is the coupling strength between the central node $0$ and the peripheral node $i$; $\alpha \in (0,\pi/2)$ is the phase shift; and $u\ge 0$ is the natural frequency detuning  (For the case $u<0$, one can obtain virtually identical results to those obtained below). Note that, the phase shift $\alpha$ is often used to model delays arising in synaptic connections  in neural networks \cite{hoppensteadt2012weakly}. 

Let $\theta=(\theta_0,\theta_1,\theta_2)^\top\in \RS^ 3$. To study the remote synchronization  in our considered network, we define a remote synchronization manifold as follows.


\begin{definition}[Remote Synchronization Manifold] \label{defi:manifold}
	The remote synchronization manifold is defined by 
	\begin{align*}
		\mathcal M:=\left\{	\theta\in \RS^3: \theta_1=\theta_2 \right\}.
	\end{align*}
\end{definition}

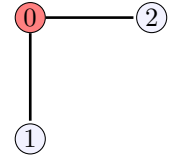
\begin{figure}[!t]
	\centering
	
	{
		\begin{tikzpicture} [->,>=stealth',shorten >=1pt,auto,node distance=1.2cm,
		main node/.style={circle,fill=blue!5,draw,minimum size=0.4cm,inner sep=0pt]},
		red node/.style={circle,fill=red!50,draw,minimum size=0.4cm,inner sep=0pt]}]
		
		\node[red node]          (0)                        {$0$};

		\node[main node]          (1) [below =of 0]          {$1$};
		\node[main node]          (2) [right =of 0]          {$2$};
		
		\tikzset{direct/.style={-,line width=1pt}}
		\tikzset{red_direct/.style={->,line width=1pt,red}}
		\path (0)     edge[direct]     node   {} (2) 
		(0)     edge[direct]     node   {} (1)    ;
		\end{tikzpicture}}	
	
	\caption{ A simple network motif: central node $0$ and peripherals $1$ and $2$.}
	\label{Fig:3Node}
\end{figure}


A solution $\theta(t)$ to \eqref{dyna:3osci} is said to be \emph{remotely synchronized} if it holds that $\theta(t)\in \mathcal M$  for all $t\ge 0$.  It is shown in \cite{nicosia2013remote} that network symmetries are critical to the occurrence of remote synchronization. In our considered network in Fig. \ref{Fig:3Node}, we say oscillators $1$ and $2$ are \textit{symmetric} if $A_1=A_2$. It can be observed that the requirement $A_1=A_2$  is necessary for the system \eqref{dyna:3osci} to have a remote synchronized solution since the equation
\begin{align*}
	\dot \theta_1 -\dot \theta_2=A_1\sin(\theta_0-\theta_1-\alpha)-A_2\sin(\theta_0-\theta_2-\alpha)=0
\end{align*}
has a solution $\theta_1=\theta_2$ \textit{only if} $A_1=A_2$. Therefore, we assume that the coupling strengths satisfy{\footnote{ This assumption requires the two coupling strengths, $A_1$ and $A_2$, to be strictly identical. This is somewhat demanding since it is not easy to be fulfilled in practical situations. Simulation results show that the phase difference between the peripheral oscillators remains bounded if $A_1$ and $A_2$ are only approximately the same, although exact phase synchronization cannot take place. It is quite interesting to study this problem in the future, though we only consider $A_1=A_2$ in this paper. }}
\begin{align}
	A_1=A_2=A. \label{assm:sym}
\end{align}
Evidently, the network in Fig. \ref{Fig:3Node} is the simplest symmetric network. In what follows, we study the exponential stability of the remote synchronization manifold $\mathcal M$ under assumption~\eqref{assm:sym}.

Define a $\delta$-neighborhood of $\mathcal M$ by $
U_{\delta} = \{\theta \in \RS^3:{\rm dist}(\theta,\mathcal M)<\delta\},$
where ${\rm dist}(\theta,\mathcal M)$ is the minimum distance from $\theta$ to a point on $\mathcal M$, that is, ${\rm dist}(\theta,\mathcal M)=\inf _{y\in \mathcal M}\|\theta-y\|$. Let us define the  exponential stability of the remote synchronization manifold $\mathcal M$.  
\begin{definition}
	For the system \eqref{dyna:3osci}, the remote synchronization manifold $\mathcal M$ is said to be exponentially stable  along the system \eqref{dyna:3osci}  if there is $\delta>0$ such that for any initial phase $\theta(0) \in\RS^ 3$ satisfying $\theta(0)\in U_{\delta}$ it holds that
	\begin{align*}
	&{\rm dist}(\theta(t),\mathcal M)=k\cdot{\rm dist}(\theta(0), \mathcal M)\cdot e^{-\lambda t},&\forall t\ge 0,
	\end{align*}
	for some $k>0$ and $\lambda>0$.
\end{definition}

In fact, remote synchronization behavior for the network~\eqref{dyna:3osci} can be categorized into two different types, depending on whether \textit{phase locking} occurs or not. Phase locking is a phenomenon where every pairwise phase difference is a constant, $\theta_i-\theta_j=c_{i,j}$, $\forall i,j$ (the phenomenon when~$c_{i,j}=0$, $\forall i,j$, is especially called phase synchronization). Phase locking is also called frequency synchronization because this is equivalent to the frequencies of all oscillators being synchronized, $\dot\theta_1=\dots=\dot\theta_n$. If the network is phase locked, it is remotely synchronized in the sense that the peripheral oscillator frequencies $\dot \theta_1$ and $\dot\theta_2$ are the same. However, the converse is not always true. In remote synchronization, the frequency of the central oscillator, $\dot{\theta}_0$, is allowed to be different from the peripheral ones, $\dot{\theta}_1,\dot{\theta}_2$. 

We will study these two categories of remote synchronization in the next two subsections, where we assume $u=0$ and $u\neq 0$, respectively, to reveal the role that the natural frequency detuning $u$ plays. The phase-locked case is relatively easy to analyze as demonstrated in the next subsection. In contrast, the analysis of the other case is technically involved, but is possible thanks to our results on partial stability established in the previous section. Although in the phase-locked case, the results on partial stability are not utilized, we analyze this property for the sake of ensuring a complete account of remote synchronization.
	
	\subsection{Natural frequency detuning $u=0$}
	In this subsection, we assume that the natural frequency detuning $u=0$. We will show that if remote synchronization occurs, it necessarily entails phase locking. A linearization method is the tool we will use  to show the stability of the remote synchronization manifold $\mathcal M$. 
	First, we observe that for any $\alpha\in(0,\pi/2)$, there always exists a remotely synchronized solution to \eqref{dyna:3osci} that is phase locked. To show this,  let $x_i:=\theta_0-\theta_i$ for $i=1,2$. The time derivative of $x_i$ is
\begin{align}\label{linearized}
	\dot x_i= \sum_{j=1}^{2}A\sin (-x_j-\alpha) -A\sin(x_i-\alpha).
\end{align}
Any remotely synchronized solution satisfies $x_1=x_2$. Solving the equation $\dot x_i=0$ with $x_1=x_2$ we obtain two isolated equilibrium points of the system \eqref{linearized} in the interval $[0,2\pi]$: 1) $x^*_1=x^*_2=c(\alpha)$; 2) $x^*_1=x^*_2=c'(\alpha)$, where
\begin{align}\label{equPoint}
	&c(\alpha)=-\arctan \left( \frac{\sin \alpha}{3\cos \alpha}\right),&c'(\alpha)=\pi+c(\alpha).
\end{align}
Note that other equilibrium points outside of $[0,2\pi]$ are equivalent to these two, and it is thus  sufficient to only consider them. Any solution satisfies $\theta_1(t)=\theta_2(t)$ and $\theta_0(t)-\theta_1(t)=c'(\alpha)$ (or $\theta_0(t)-\theta_1(t)=c(\alpha)$) is a phase-locked and remotely synchronized solution. To capture this type of remote synchronization, we define $\mathcal M_1:=\{\theta\in \mathcal M: \theta_0-\theta_1=c(\alpha)\}$, $\mathcal M'_1:=\{\theta\in \mathcal M: \theta_0-\theta_1=c'(\alpha)\}$, and refer to them as the \textit{phase-locked} remote synchronization manifolds. It is not hard to see that they are two positively invariant manifolds of the system \eqref{dyna:3osci}. We show in the following theorem that $\mathcal M'_1$ is always unstable, and the phase shift $\alpha$ plays an essential role in determining the stability of $\mathcal M_1$. The proof using the linearization method can be found in Appendix \ref{proof:linear}.

\begin{theorem}\label{theo:inear}
	Assume that \eqref{assm:sym} is satisfied. For any nonzero $A$, the following statements hold:
	\begin{enumerate}
		\item  if $\alpha < \arctan \left(\sqrt{3} \right)$, there exists a unique exponentially stable remote synchronization manifold in $\mathcal M$, that is $\mathcal M_1$;
		\item if $\alpha >\arctan \left(\sqrt{3} \right)$, there does not exist an exponentially stable remote synchronization manifold in $\mathcal M$ .
	\end{enumerate}	
\end{theorem}

Consistent with the findings in \cite{vicente2008dynamical} and \cite{fischer2006zero}, remote synchronization emerges thanks to the central mediating oscillator, and it is exponentially stable  for a wide range of phase shift, i.e., $\alpha\in (0,\arctan (\sqrt{3} ))$. Nevertheless, an even larger phase shift $\alpha$ out of this range can destabilize  the remote synchronization. In the next subsection, we detune the natural frequency of the central oscillator by letting $u\neq 0$, which is similar to the introduction of parameter impurity in \cite{banerjee2012enhancing}, and show how a sufficiently large natural frequency detuning can lead to robust remote synchronization that is exponential stable for any phase shift $\alpha\in (0,\pi/2)$.
	

\subsection{ Natural frequency detuning $u\neq 0$}

In this subsection, we consider the case when the natural frequency detuning $u> 0$, and show how it can give rise to robust remote synchronization. 

Note that if $u>3A$, there does not exists a phase-locked solution to \eqref{dyna:3osci}. This is because the equations $\dot x_i= u+\sum_{j=1}^{2}A\sin (-x_j-\alpha) -A\sin(x_i-\alpha)=0,i=1,2,$ do not have a solution. Nevertheless, the remote synchronization can still be exponentially stable for a sufficiently large control input $u$. In other words, the natural frequency detuning can actually stabilize the remote synchronization, although it makes phase locking impossible. In fact, the remote synchronization manifold $\mathcal M$ as a whole becomes exponentially stable for any $\alpha$ with this control input. 
The following is the main result of this section.

\begin{theorem}\label{theo:stabilization}
	There is a positive constant $\rho>3A$  such that for any $u$ satisfying $u>\rho$, the remote synchronization manifold $\mathcal M$ is exponentially stable for the system \eqref{dyna:3osci} for any phase shift $\alpha\in (0,\pi/2)$. 
\end{theorem}

The proof is technically involved and is based on the results on partial stability established in the previous section. Before providing the proof, we first define some variables, and  associate the remote synchronization manifold $\mathcal M$ with an equivalent set defined on the new variables. We then prove that this set is exponentially stable.

 Let us define~$z_1$ and~$z_2$ by
	\begin{subequations}\label{nota:z}
		\begin{align}
			&z_1:=\frac{1}{2}\sum_{j=1}^{2}\cos (\theta_0-\theta_j), \\
			&z_2:=\frac{1}{2}\sum_{j=1}^{2}\sin (\theta_0-\theta_j).
		\end{align}
\end{subequations}
Then, it is clear to see  $z_1,z_2\in \R$ satisfy $|z_1|\le 1$ and $|z_2|\le 1$.
Note that for any initial condition $\theta(0)\in \RS^3$, the unique solution $\theta(t)$ to \eqref{dyna:3osci} exists for all $t\ge 0$. As a consequence, $z_1(t)$ and $z_2(t)$ exist for all $t\ge 0$.
We then define the following unit circle by using $z_1$ and $z_2$:
\begin{align}
	\mathcal L:=\Big\{z\in \mathbb{R}^2: z_1^ 2 + z_2^ 2=1\Big\}, \label{limCyc:formal}
\end{align}
where $z=(z_1,z_2)^\top$. In fact, this set~$\mathcal L$ has a strong relation with remote synchronization as follows.
\begin{proposition}\label{equi:lim:solu}
	Let $\mathcal M$ and $\mathcal L$ be defined in Definition \ref{defi:manifold} and \eqref{limCyc:formal}, respectively.  The following two statements are equivalent:
	\begin{enumerate}
		\item $\theta$ belongs to the remote synchronization manifold $\mathcal M$. 
		\item $z=(z_1,z_2)^\top$ belongs to the set $\mathcal L$.
	\end{enumerate}
	
\end{proposition}
\begin{IEEEproof}
	From \eqref{nota:z}, there holds that
	\begin{align}
		z_1^ 2 + &z_2^ 2= \nonumber\\
		& \frac{1}{4}\Big( \sum_{j=1}^{2}\cos (\theta_0-\theta_j) \Big)^2 + \frac{1}{4}\Big(\sum_{j=1}^{2}\sin (\theta_0-\theta_j) \Big)^2. \label{limi:1}
	\end{align}
	The right-hand side of the equality \eqref{limi:1} can be simplified to
	\begin{align}
		&\frac{1}{4}\Big( \sum_{j=1}^{2}\cos (\theta_0-\theta_j) \Big)^2 + \frac{1}{4}\Big(\sum_{j=1}^{2}\sin (\theta_0-\theta_j) \Big)^2 \nonumber \\
		&=\frac{1}{4}\sum_{j=1}^{2}\left(\cos^ 2 (\theta_0-\theta_j)+\sin ^ 2 (\theta_0-\theta_j)\right) \nonumber\\
		&+\frac{2}{4}\cos(\theta_0-\theta_1)\cos(\theta_0-\theta_2) 
		+\frac{2}{4}\sin(\theta_0-\theta_1)\sin(\theta_0-\theta_2)\nonumber \\
		&=\frac{1}{2}+\frac{1}{2}\cos(\theta_1-\theta_2), \label{simp:r}
	\end{align}
	where the last equality has used the trigonometric identity $\cos a \cos b+\sin a \sin b=\cos(a-b)$. 	
	We first prove that $1)$ implies $2)$. If $\theta\in \mathcal M$, one obtains $\theta_1=\theta_2$ from the  definition of the remote synchronization manifold $\mathcal M$. It follows from \eqref{simp:r} that the right-hand side of \eqref{limi:1} equals $1$. 	
	We now prove that $2)$ implies $1)$. If $z\in \mathcal L$, from \eqref{simp:r} we obtain $1/2+1/2\cos(\theta_1-\theta_2)=1$,  which means that  $\cos(\theta_1-\theta_2)=1$. This, in turn, proves that $\theta\in \mathcal M$. The proof is complete. 
\end{IEEEproof}

Proposition \ref{equi:lim:solu} provides us an alternative way to study remote synchronization. 
Any pair of $(z_1,z_2)$ belongs to $\mathcal L$ if and only if the corresponding $\theta \in \mathbb{R}^{3}$ is included in the remote synchronization manifold $\mathcal M$. Further, if $\theta_1(0)=\theta_2(0)$, it can be seen from \eqref{dyna:3osci} that $\theta_1(t)=\theta_2(t)$ for all $t\ge 0$. In other words, $z(0)\in \mathcal L$ implies that $z(t)\in \mathcal L$ for all $t\ge 0$, which means that the set  $\mathcal L$ is a positively invariant set of the system \eqref{dyna:3osci}. 
To show the exponential stability of the remote synchronization manifold, it suffices to show the positively invariant set $\mathcal L$ is exponentially stable along the system \eqref{dyna:3osci} using the distance ${\rm dist}(z,\mathcal L)=\inf _{y\in \mathcal L}\|z-y\|$.

To proceed with the analysis, we represent $z_1$ and $z_2$ in the polar coordinates
\begin{align}
	&z_1=r\cos   \zeta, \label{polar:cos}\\
	&z_2=r \sin   \zeta,\label{polar:sin}
\end{align}
where with~\eqref{nota:z},
\begin{align}
	r&:=\frac{1}{2}\sqrt{\Big( \sum_{j=1}^{2}\cos (\theta_0-\theta_j) \Big)^2 + \Big(\sum_{j=1}^{2}\sin (\theta_0-\theta_j) \Big)^2}, \label{expre:r}\\
	\zeta&:= \arctan \left(\frac{\sum_{j=1}^{2}\sin (\theta_0-\theta_j)}{\sum_{j=1}^{2}\cos (\theta_0-\theta_j)}\right).\label{expre:phi}
\end{align}
It follows from \eqref{polar:cos} and \eqref{polar:sin} that $z_1^2(t)+z_2^2(t)=r^2$. Thus, the distance from $z(t)$ to the circle $\mathcal L$, denoted by~$\mu(t)$ is 
\begin{align}\label{distance}
	\mu(t):={\dist}(z(t),\mathcal L)=1-r(t).
\end{align}
	In fact, the dynamics of~$\mu(t)$ and $\zeta(t)$ can be described by
	\begin{subequations}\label{expres:r:3:phi}
	\begin{align}
		&\frac{d \mu(t)}{d t}= -A(1-(1-\mu)^2)\cos( \zeta-\alpha), \label{expres:r:3}\\
		&\frac{d \zeta(t)}{dt}=u-A(1-\mu)\big (2\sin(\zeta+\alpha)+\sin(\zeta-\alpha) \big), \label{exre:dphi_dt:1}
	\end{align}
\end{subequations}
	where~$\mu(t) \in [0,1]$ and~$\zeta (t) \in \R$. The proof is given by Proposition~\ref{dynamics:prop} in Appendix~\ref{proposition:04}. Therefore, remote synchronization analysis reduces to partial stability analysis of~$\mu(t)$. We are now ready to provide the proof of Theorem~\ref{theo:stabilization} based on the results of partial stability obtained in the previous section.

\begin{IEEEproof}[Proof of Theorem \ref{theo:stabilization}]
	As we have shown, in order to prove the exponential stability of $\mathcal M$, it is sufficient to prove the set $\mathcal L$ is partially exponentially stable along the system \eqref{dyna:3osci}. In other words, we show that~$\mu=0$ of the system \eqref{expres:r:3:phi} is exponentially stable uniformly in $ \zeta$ based on Corollary~\ref{Coro:2:aver}. To this end, we show that the system satisfies the conditions in  Corollary~\ref{Coro:2:aver}.
	
	First, we confirm the requirements for the system~\eqref{simplified} taking $\mu =x_1$ and $\zeta =z$. These are simply that ~$\mu=0$ is a partial equilibrium of~\eqref{expres:r:3}, and the system is $2\pi$-periodic in $\zeta$, which one can check immediately.  Next, from the assumption $u>\rho>3A$, there exists $\vartheta>0$ such that  ${d \zeta(t)}/{dt}\ge \vartheta$ for any $\mu$ and $\zeta$, which means that assumption~\eqref{f3:bound} is satisfied. 
		
		Next, we compute the equivalent of~\eqref{simplified_reduced:1} by applying the change of time-axis, $t \mapsto \zeta$, and then obtain the equivalent of  its averaged system~\eqref{simplified:aver}. The derivative of $\mu$ with respect to $\zeta$ can be computed as 
	\begin{align}
		\frac{d \mu}{d \zeta}
		=\frac{d r}{d t} /\frac{d \zeta}{d t}=\varepsilon f(\mu,\zeta) \label{dmu:dphi}
	\end{align}
	where, corresponding to $h_1$, 
	\begin{align*}
		f(\mu,\zeta)=\frac{-A(2-\mu)\mu\cos(\zeta-\alpha)}{1-\frac{A}{u}(1-\mu) \Big(2\sin(\zeta+\alpha)-\sin(\zeta-\alpha)\Big)},
	\end{align*}
	and $\varepsilon=1/u$. Note that for any given $u$, there is $L>0$ such that the equivalent of~\eqref{simplified:aver:bound} holds. Then, the associated averaged system is
	\begin{align}
		\dot {\hat \mu} =f_\av (\hat \mu):=\int_{0}^{2\pi}f(\hat \mu,\tau)d\tau,\label{aver:mu}
	\end{align}
	where
	\begin{align*}
		& f_\av (\hat\mu)=\int_{0}^{2\pi}f(\hat\mu,\tau)d\tau=\frac{4\pi (2 - \hat\mu )\hat\mu }{(1 - \hat\mu)(5 + 4\cos 2\alpha )}\cdot g(\hat\mu),\\
		&g(\hat \mu)=\left( \frac{1}{u}-\frac{1}{\sqrt{u^2 - 5 A^2 (1-\hat\mu)^2 -4A^2(1-\hat\mu)^2 \cos 2\alpha}} \right).
	\end{align*}
	
	According to Corollary~\ref{Coro:2:aver}, it remains to check the exponential stability of the averaged system. By the assumption $u> 3A$, it follows that  $g(\hat\mu)<0$ and thus $f_\av(\hat\mu)<0$ for any $0<\hat\mu<1$. Moreover, for any $\hat\mu$ satisfying $0<\hat\mu<\xi<1$, it holds that 
	\begin{align}
		f_\av (\hat\mu)<-c \hat\mu,
	\end{align}
	where the constant $c$ is given by
	\begin{align*}
		c=\frac{4 \pi }{9}\left(\frac{1}{\sqrt{u^2 - 9 A^2 (1-\xi)^2}} - \frac{1}{u}\right).
	\end{align*}
	Choose $V(\hat \mu)=\hat \mu^2$ as a Lyapunov candidate, and it is easy to see $\dot V {\le} -c\hat \mu^2$, which implies that $\hat \mu =0$ is exponentially stable along the averaged system \eqref{aver:mu} for any $u>3A$. 
	According to  Corollary 1, there exists $\varepsilon^*>0$ such that if $\varepsilon<\varepsilon^*$, the system \eqref{expres:r:3:phi} is partially exponentially stable with respect to $\mu$. As $\varepsilon=1/u$, it is equivalent to saying that there exists $\rho>3A$ such that if the input $u>\rho$, the system \eqref{expres:r:3:phi} is partially exponentially stable respect to $\mu$. Thus, the remote synchronization manifold $\mathcal M$ is exponential stable for any phase shift $\alpha$. 
\end{IEEEproof}

Consistent with the experimental findings in \cite{banerjee2012enhancing,gambuzza2016inhomogeneity}, Theorem \ref{theo:stabilization} rigorously shows that by detuning the natural frequency one is able to stabilize the remote synchronization manifold even when the phase shift is quite large.  Interestingly, the central oscillator has a different frequency $\dot \theta_0$ from the peripheral ones $\dot \theta_1$ and $\dot \theta_2$ when remote synchronization occurs under the assumption $u>\rho>3A$.

In fact, what we have proven in Theorem \ref{theo:stabilization} is that $\mathcal L$ is an exponentially stable limit cycle. We now show that it is also periodic. 
It can be observed that the set $\mathcal L$ defined in \eqref{limCyc:formal} is compact. From \eqref{polar:cos} and \eqref{polar:sin}, the time derivatives of $z_1$ and $z_2$ are $\dot z_1=\dot r \cos \zeta- r\sin \zeta \cdot \dot \zeta$ and $\dot z_2=\dot r\sin \zeta+r\cos \zeta \cdot \dot \zeta$, respectively. Since $\dot r=-\dot \mu$ and $r=1-\mu$, it follows that 
\begin{subequations}\label{dynamics:z_1+z_2}
	\begin{align}
	&\dot z_1=-\dot \mu \cos \zeta- (1-\mu)\sin \zeta \cdot \dot \zeta,\\
	&\dot z_2=-\dot \mu \sin \zeta+(1-\mu)\cos \zeta \cdot \dot \zeta.
	\end{align}
\end{subequations}
From \eqref{distance},  one knows that $\mu=0$ and $\dot \mu=0$ if $\col(z_1,z_2)\in \mathcal L$, in which case the dynamics \eqref{dynamics:z_1+z_2} become $\dot z_1=-\dot \zeta \sin \zeta, \dot z_2= \dot \zeta \cos \zeta$. One can observe that this system has no equilibrium in $\mathcal L$ since it holds that $\dot \zeta >0$ from \eqref{exre:dphi_dt:1}. It is implied by Poincar\'{e}-Bendixson Theorem \cite[Theorem 9.3]{terrell2009stability} that $\mathcal L$ is actually a \textit{periodic orbit}. In other words, $\mathcal L$ is an exponentially stable periodic orbit.


Let $v_1=\dot \theta_1+\dot\theta_2$ and $v_2=\dot \theta_0$, and then we can rewrite the set \eqref{limCyc:formal} into 
\begin{align}\label{limitcycle:frequencies}
	\mathcal C:&=\Big\{ (v_1,v_2)^\top\in\R^2:\nonumber\\
	&\frac{(v_1+v_2-3\omega-u)^2}{16A^ 2\sin^ 2\alpha}+\frac{(v_1-v_2-\omega+u)^2}{16A^ 2\cos^ 2\alpha}=1\Big\},
\end{align}
which is also a limit cycle for the variables $v_1$ and $v_2$. Note that $v_1$ is the sum of the peripheral oscillators' frequencies. One can say the remote synchronization is reached if and only if the frequencies $v_1$ and $v_2$ reach the limit cycle $\mathcal C$. Theorem \ref{theo:stabilization} also implies the exponential stability of the limit cycle $\mathcal C$, and furthermore, $\mathcal C$ is also periodic.

\section{Concluding Remarks}\label{sec:conclusion}

Using periodic averaging methods, we have obtained some criteria for partial exponential stability of a type of slow-fast nonlinear systems in this paper.  We have associated the original system with an averaged one by averaging over the fast varying variable. We have shown that the partial exponential stability of the averaged system implies that of the original one.  As some intermediate results, we have also obtained: 1) a converse Lyapunov theorem; and 2) some perturbation theorems that are the first known ones for partially exponentially stable systems. We have applied our results to stabilization of remote synchronization in a network of Kuramoto oscillators with a phase shift, showing how our results can be used to prove stability of limit cycles. In the future, we are interested in developing new criteria for partial asymptotic stability of nonlinear systems using averaging techniques. Moreover, it is interesting to study stability of remote synchronization in more complex networks.

\section{Appendix}

\subsection{Proof of Theorem \ref{Theo:converse}}\label{proof_Theo:2}
Before proving Theorem \ref{Theo:converse}, we first present an intermediate result. 
\begin{proposition}\label{bound:partial:f_y}
	Consider a continuously differentiable function $h:\R^{n_1}\times \R^{n_2}\to \R^{n_1}$, which satisfies $h(0,v)=0$ for any $v\in \R^{n_2}$. Suppose that there exists a connected set $\mathcal D\subset \R^{n_1}$ containing the origin $x=0$ such that 
	\begin{align*}
		&\left\| \frac{\partial h}{\partial w}(w,v)\right\|\le l_1, &\forall w \in \mathcal D ,v \in \R^{n_2},
	\end{align*}
	for a positive constant $l_1$.
	Then, there exists $l_2>0$ such that 
	\begin{align*}
		&\left\| \frac{\partial h}{\partial v}(w,v)\right\|\le l_2 \|w\|, &\forall w \in \mathcal D, v \in \R^{n_2}.
	\end{align*}
\end{proposition}
\begin{IEEEproof}
	Due to the continuous differentiability of $h$, it follows from the mean value theorem that for any $w\in \mathcal D$, and $v,\delta\in \R^{n_2}$, there exists $0\le \lambda \le 1$ such that
	\begin{align*}
		h_i(w,v+\delta) -h_i(w,v)=\frac{\partial h_i}{\partial v}(w,v+\lambda \delta)\cdot \delta.
	\end{align*}
	Since $\|\partial h/\partial w\|\le l_1$ and $h(0,v)=0, \forall v$, we have 
	\begin{align}
		&\|h_i(w,v+\delta )-h_i(w,v)\|\nonumber\\
		&\le \|h_i(w,v+\delta)-h_i(0,v+\delta)\|+ \|h_i(w,v)-h_i(0,v)\| \nonumber\\
		&\le 2l_1\|w\|, \label{ine:h:i}
	\end{align}
	where the last inequality has used the fact that $\|\partial h_i/\partial w\|\le \|\partial h/\partial w\|\le l_1$. We then observe that the inequality
	\begin{align*}
		\left\| \frac{\partial h_i}{\partial v}(w,v+\lambda  \delta)\cdot \delta \right\| \le 2l_1\|w\|
	\end{align*} 
	holds for any  $w\in \mathcal D$, and $v,\delta\in \R^{n_2}$, which implies that for any $i$ there exists $l'_1>0$ such that $\|\partial h_i/\partial v\|\le l'_1\|w\|$. As a consequence, there is $l_2>0$ so that $\|\partial h/\partial v\|\le l_2\|w\|$. 	
\end{IEEEproof}

We are now ready to prove Theorem \ref{Theo:converse}. 
\begin{IEEEproof}[Proof of Theorem \ref{Theo:converse}]
	Let $\phi_1(\tau;w,v,z)$ and $\phi_2(\tau;w,v,z)$ denote the solution to the system \eqref{side:1} and \eqref{side:2} that starts at $(w,v,z)$, respectively; note that $\phi_1(z;w,v,z)=x$ and $\phi_2(z;w,v,z)=y$.  Let 
	\begin{align}\label{Lyapun:expo}
		V(w,v,z)=\int_{z}^{z+\delta} \|\phi_1(\tau;w,v,z)\|^2_2 d\tau. 
	\end{align}
	Following similar steps as those in Theorem 4.14 of \cite{khalil2002nonlinear} and Theorem 4.4 of \cite{haddad2011nonlinear}, one can show that 
	\begin{align*}
		\frac{1}{2L_1} (1-e^{-2L_1\delta}) \|w\|_2^2 \le V(w,v,z) \le\frac{k^2}{2\lambda}(1-e^{-2\lambda \delta}) \|w\|_2^2,
	\end{align*}
	and $ \dot V(w,v,z)=-(1-k^2e^{-2\lambda \delta})\|w\|_2^2$, which proves the inequalities \eqref{conver:bounds} and \eqref{conver:decay}. Note that $\dot V$ denotes a total derivative along trajectories with the independent variable $z$ rather than $t$.

	We next prove the inequalities \eqref{parti-V:bounds} and \eqref{parti-V:bounds:y}. Let $\phi=\col(\phi_1,\phi_2), \varphi=\col(\varphi_1,\varphi_2)$, and for $k=1,2$, denote
	\begin{align*}
		\phi_{k,w}(\tau;w,v,z)= \frac{\partial}{\partial w}\phi_{k}(\tau;w,v,z).
	\end{align*}
	and $\phi'_w=\partial \phi/\partial w$. Note that $\phi_1(\tau;w,v,z)$ and $\phi_2(\tau;w,v,z)$ satisfy 
	\begin{align*}
		&\phi_1(\tau;w,v,z)=w+\int_{z}^{\tau} \varphi_1(\phi_1(s;w,v,z),\phi_2(s;w,v,z))ds,\\
		&\phi_2(\tau;w,v,z)=v+\int_{z}^{\tau} \varphi_2(\phi_2(s;w,v,z),\phi_2(s;w,v,z))ds.
	\end{align*}
	Then, the partial derivative  $\phi'_w$ is
	\begin{align}\label{equality:phi_x}
		\phi'_w(\tau;w,v,z)=\left[ {\begin{array}{*{20}{c}}
				I&0\\
				0&0
		\end{array}} \right] + \int_z^\tau  {\frac{\partial \varphi}{\partial \phi}\phi'_x(s;w,v,z)} ds.
	\end{align}
	Recall that  $\varphi_1$ and $\varphi_2$ are continuously differentiable, and satisfy $\varphi_1(0,v,z)=0$ and $\varphi_2(0,v,z)=0$ for any $v$ and $z$, and $\|\partial \varphi_1/\partial w\|_2\le L_1$ and $\|\partial \varphi_2/\partial w\|_2\le L_2$ for any $w\in \mathcal B_\delta$ and $v\in \R^{m},t\in \R$. It then follows from Proposition \ref{bound:partial:f_y} that $\|\partial \varphi_1/\partial v\|_2\le L'_1$ and $\|\partial \varphi_2/\partial v\|_2\le L'_2$ for some positive constants $L'_1$ and $L'_2$, since $\|w\|$ is upper bounded by some constant in a compact set $\mathcal B_\delta$. Equivalently, there exists $L>0$ such that 
	\begin{align}\label{bound:partialf_X}
		&\left\| \frac{\partial \varphi}{\partial X}\right\|_2\le L, &\forall X\in \mathcal B_\delta \times \R^{m},
	\end{align}
	where $X=\col(w,v)$. 
	Consequently, $\|\partial \varphi/\partial \phi\|_2\le L$, and it then follows from \eqref{equality:phi_x} that 
	\begin{align*}
		\|\phi'_w(\tau;w,v,z)\|_2\le 1 + L\int_{z}^\tau  \|\phi'_w(s;,w,v,z)\|_2ds,
	\end{align*}
	which implies that $\|\phi'_{w}(\tau;w,v,z)\|_2 \le e^{L(\tau-z)}$ by Gr{\"o}nwall's lemma \cite[Lemma 2.2]{haddad2011nonlinear}. Since $\|\phi_{1,w}\|_2 \le \|\phi_{1,w},\phi_{2,w}\|_2=\|\phi'_{w}(\tau;w,v,z)\|_2$, it holds that 
	\begin{align}\label{bound:phi:x}
		\|\phi_{1,w}(\tau;w,v,z)\|_2 \le  e^{L(\tau-z)}.	
	\end{align}
	From \eqref{expo:decay}, it holds that $\|\phi_1(\tau;w,v,z)\|\le K\|w\|_2e^{-\lambda(\tau-z)}$. Then, the partial derivative $\partial V/\partial w$ satisfies
	\begin{align*} 
		\left\|\frac{\partial V}{\partial w}\right\|_2& =\left\|\int_{z}^{ z+\delta} 2 \phi_1^\top(\tau;w,v,z) \phi_{1,w}(\tau;w,v,z)d\tau\right\|_2\\
		&\le \int_{z}^{z+\delta} 2\left\| \phi_1(\tau;w,v,z) \right\|\cdot \|\phi_{1,w}(\tau;w,v,z)\|_2 d\tau\\
		&\le  \int_{z}^{z+\delta} 2k \|w\|_2 e^{-\lambda (\tau-z)} e^{L(\tau-z)}d\tau =c_4 \|w\|_2,
	\end{align*}
	with $c_4=2k(1-e^{-(\lambda-L)\delta})/(\lambda-L)$, which proves the inequality \eqref{parti-V:bounds}. Note that the case when $\lambda=L$ can always be ruled out since one can replace $L$ with $L+\sigma$ for any $\sigma>0$ if that happens.

	Following a similar line, one can show there exists $c_5>0$ such that \eqref{parti-V:bounds:y} are satisfied, which completes the proof.
\end{IEEEproof}

\subsection{Proof of Theorem \ref{Theo:asymBeha}}\label{proof:the03}
\begin{IEEEproof}
	From the assumption for the nominal system \eqref{side:main}, there is a Lyapunov function~$V$ satisfying all conditions in Theorem~\ref{Theo:converse}. Here, we use this $V$ to estimate the convergence speed of the perturbed system~\eqref{side:perturbed}.
	
	For any $w_p\in \mathcal B_\delta$ and $v_p\in \R^m,z\in \R$, the total derivative of the Lyapunov function $V(w_p,v_p,z)$ along the trajectories of the perturbed system \eqref{side:perturbed} satisfies 
	\begin{align*}
		\frac{dV}{dz}=&{\frac{\partial V}{\partial z}} +\frac{\partial V}{\partial w_p}\left( {\varphi_1(w_p,v_p,z)}+g_1(w_p,v_p,z)\right)\\
		&+\frac{\partial V}{\partial v_p}\left({ \varphi_2(w_p,v_p,z)}+g_2(w_p,v_p,z)\right)\\
		\le&-c_3\|w_p\|^2+\left(c_4 \gamma_1 (z)+c_5 \gamma_2 (z)\right)\|w_p\|^2\\
		&+\left( c_4 \psi_1(z)+c_5\psi_2(z) \right)\|w_p\|,
	\end{align*}
	where the last inequality follows from the inequalities \eqref{conver:decay}, \eqref{parti-V:bounds}, and \eqref{parti-V:bounds:y} together with the bounds \eqref{pert:1} and \eqref{pert:2}. For simplicity, denote $\gamma(z)=c_4 \gamma_1 (z)+c_5 \gamma_2 (z)$ and $\psi(z)=c_4 \psi_1(z)+c_5\psi_2(z)$. By the inequality \eqref{conver:bounds} one can obtain an upper bound for $\dot V$ given by 
	\begin{align*}
		\frac{dV}{dz} \le -\left[\frac{c_3}{c_2}- \frac{1}{c_1}\gamma(z)\right]V+\frac{1}{\sqrt{c_1}}\psi(z)\sqrt{V}.
	\end{align*}
	Let $W(w_p,v_p,z)=\sqrt{V(w_p,v_p,z)}$, and when $V\neq 0$  the total derivative of $W$ satisfies
	\begin{align}
		\frac{d W}{dz} = \frac{d V/dz}{2\sqrt{V}}\le -\frac{1}{2}\left[\frac{c_3}{c_2}- \frac{1}{c_1}\gamma(z)\right]W+\frac{1}{2\sqrt{c_1}}\psi(z). \label{deri:W}
	\end{align}
	When $V=0$, following the same idea as the proof of Lemma 9.4 in \cite{khalil2002nonlinear} one can show that the Dini derivative of $W$ satisfies
	\begin{align*}
		&D^+W\le \frac{1}{2\sqrt{c_1}}\psi(z),
	\end{align*}
	which implies that $D^+W$ satisfies \eqref{deri:W} for all values of  $V$. 
	Using the comparison lemma \cite[Lemma 3.4]{khalil2002nonlinear}, one can show that $W(z)$ satisfies the following inequality
	\begin{align}\label{expre:W}
		W(z)\le \Phi(z,z_0) W(z_0)+\frac{1}{2\sqrt{c_1}}\int_{z_0}^{z}\Phi(z,\tau)\psi(\tau)d\tau,
	\end{align}
	where the transition function is 
	\begin{align}\label{expres:Phi}
		\Phi(z,z_0)=\exp\left( - \frac{c_3}{2c_2}(z-z_0)+\frac{1}{2c_1}\int_{z_0}^{z}\gamma(\tau)d\tau\right).
	\end{align}
	Substituting \eqref{Theo3:Asp1} and \eqref{Theo3:Asp2} into \eqref{expres:Phi}, we have
	\begin{align}\label{expres:Phi_bound}
		\Phi(z,z_0)\le k_2 e^{-k_1(z-z_0) },
	\end{align}
	with $k_1$ and $k_2$ given by \eqref{expre:k_1,k_2}. 
	
	From the inequality \eqref{conver:bounds}, we obtain $\sqrt{c_1}\|w_p\|\le W \le \sqrt{c_2}\|w_p\|$. Then, for any $z\ge z_0$ such that $\|w_p(z)\|\in \mathcal B_\delta$, it follows from the inequalities \eqref{expre:W} and~\eqref{expres:Phi_bound} that 
	\begin{align}\label{bound:x(t)}
		\|w_p(z)\| \le k_2 \sqrt{\frac{c_2}{c_1}}&\|w_p(z_0)\|e^{-k_1(z-z_0) } \nonumber\\ &+\frac{k_2}{2c_1}\int_{z_0}^{z}e^{-k_1(z-\tau) }\psi(\tau)d\tau,
	\end{align}
	for any $z\ge z_0$ such that $\|w_p(z)\|\in \mathcal B_\delta$. Under the assumption \eqref{Theo3:Asp3}, if the initial condition satisfies \eqref{init:x0}, it then holds that 
	\begin{align*}
		\|w_p(z)\| < \delta e^{-k_1(z-z_0)}+\delta (1-e^{-k_1(z-z_0)})=\delta, \forall z \ge z_0,
	\end{align*}
	which ensures that the inequality \eqref{bound:x(t)} holds for any $z\ge z_0$. The proof is complete. 
\end{IEEEproof}

\subsection{Proposition \ref{u:buound}}\label{proposition:02}
\begin{proposition}\label{u:buound}
	Consider the function~$u(w_p,v_p,z)$ defined in~\eqref{defi:u}.
	For any $w_p\in \mathcal B_\delta$,~$v_p\in \R^{m},$~$z\in \R$, $\|u(w_p,v_p,z)\|$, $\|\partial u/\partial w_p\|$, and $\|\partial u/\partial v_p\|$ are all bounded. 
\end{proposition}
\begin{IEEEproof}
	First, we prove that $\|u(w_p,v_p,z)\|$ is bounded. 
	One can observe that $u(w_p,v_p,{z})$ is $T$-periodic in ${z}$ since $\Delta(w_p,v_p,{ z})$ is. For any ${ z}\ge 0$, there exists a nonnegative integer $N_1$ and ${z}'$ satisfying $0\le {z}'< T$. such that ${z}=N_1T+{z}'$.  Then, using \eqref{Delta:0mean} we have
	\begin{align*}
		&\int_{0}^{{z}} \Delta(w_p,v_p,\tau)d\tau\\
		&=\int_{0}^{N_1T} \Delta(w_p,v_p,\tau)d\tau+\int_{0}^{{z}'} \Delta(w_p,v_p,\tau)d\tau,\\
		&=\int_{0}^{{z}'} \Delta(w_p,v_p,\tau)d\tau.
	\end{align*}
	
	Next, the partial derivative of $\Delta$ with respect to $w$ satisfies
	\begin{align*}
		&\left\| \frac{\partial \Delta}{\partial w_p}(w_p,v_p,{z}) \right\|\\
		&=\left\|\frac{\partial h_1}{\partial w_p}(w_p,v_p,{ z})- \frac{1}{T} \int_{0}^{T}\frac{\partial h_1}{\partial w_p}(w_p,v_p,\tau)d\tau\right\| \le 2L_1,
	\end{align*}
	where the triangle inequality and \eqref{aver:bound} have been used. Using this inequality and $\Delta(0,v_p,{ z})=0$ in \eqref{defi:u} yields
	\begin{align}
		\|u(w_p,v_p,{z})\|&\le \int_{0}^{{ z}'} \|\Delta(w_p,v_p,\tau)-\Delta(0,v_p,\tau)\|d\tau \nonumber\\
		& \le 2z'L_1\|w_p\| \le 2TL_1\|w_p\|. \label{bound:u}
	\end{align}
	For any $w_p\in \mathcal B_\delta$ and $v_p\in \R^{m},{z}\in \R$, it is now clear that $\|u(w_p,v_p,{ z})\|\le 2TL_1 \delta$.

	Second, we prove $\|{\partial u}/{\partial w_p}\|$ is bounded.
	Since ${\partial u}/{\partial w_p}$ is $T$-periodic in ${ z}$ and satisfies $\int_{0}^{T}\frac{\partial u}{\partial w_p}(w_p,v_p,\tau)d\tau=0$, for any ${z}\ge 0$, there exists a nonnegative integer $N_2$ and ${z}''$ satisfying $0\le {z}''\le T$ such that ${z}=N_2T+{z}''$, implying then that 
	\begin{align*}
		\frac{\partial u}{\partial w_p}(w_p,v_p,{ z})=\int_{0}^{{ z}''} \frac{\partial \Delta}{\partial w_p}(w_p,v_p,\tau)d\tau.
	\end{align*}
	In turn, this implies that for any $w_p\in \mathcal B_\delta$, $v_p\in \R^{m}$, and ${ z}\in \R$,
	\begin{align*}
		\left\|\frac{\partial u}{\partial w_p}(w_p,v_p,{ z})\right\| &\le \int_{0}^{{ z}'' } \left\| \frac{\partial \Delta}{\partial w_p }(w_p,v_p,\tau) \right\| d\tau \\ &\le 2 { z}'' L_1\le 2TL_1.
	\end{align*}
	
	Finally, we prove $\|{\partial u}/{\partial v_p}\|$ is bounded. 
	Since $h_1(0,v_p,{ z})=0$ for any $v_p$ and ${ z}$, from Proposition \ref{bound:partial:f_y}, there exists $L'_1>0$ such that for any $w_p\in \mathcal B_\delta$, and $v_p\in \R^{m},{ z}\in \R$,
	\begin{align}
		\left \|\frac{\partial h_1}{\partial v_p} (w_p,v_p,{ z})\right\|\le L'_1 \|w_p\|. 
	\end{align}
	Then, the partial derivative of $\Delta$ with respect to $v_p$ satisfies
	\begin{align*}
		&\left\| \frac{\partial \Delta}{\partial v_p}(w_p,v_p,{ z}) \right\|\\
		&=\left\|\frac{\partial h_1}{\partial v_p}(w_p,v_p,{ z})- \frac{1}{T} \int_{0}^{T}\frac{\partial h_1}{\partial v_p}(w_p,v_p,\tau)d\tau\right\| \\
		& \le 2 L'_1 \|w_p\|,
	\end{align*}
	which implies that $\left\|\frac{\partial u}{\partial v_p}(w_p,v_p,{ z})\right\| \le 2TL'_1  \|w_p\|\le 2TL'_1 \delta$ for any $w_p\in \mathcal B_\delta$. 
\end{IEEEproof}

\subsection{Proof of Theorem \ref{theo:inear}}\label{proof:linear}
\begin{IEEEproof}
	We  prove this theorem by two steps. We first demonstrate that $\mathcal M_1$ and $\mathcal M'_1$ are the only two positively invariant manifolds in $\mathcal M$ for any $\alpha$ by proving that starting from any point in $\mathcal M/\mathcal M'_1$, the solution to \eqref{dyna:3osci},  $\theta(t)$, converges to $\mathcal M_1$ asymptotically. Then, we investigate the stability of $\mathcal M_1$ and $\mathcal M'_1$ under different assumptions of $\alpha$.
	
	We start with the first step.   When $\theta\in\mathcal M$, there holds that $x_1=x_2$. Then, the dynamics of $x_1$ and $x_2$ are described by
	\begin{align}\label{dynamics:M}
	\dot x_i= -2A\sin (x_i+\alpha) -A\sin(x_i-\alpha).
	\end{align}
	Note that $x_2$ has the same dynamics of $x_1$, and it is thus sufficient to only investigate the asymptotic behavior of $x_1$.
	For any initial condition $\theta(0)\in \mathcal M/\mathcal M'_1$, there hold that $\theta_1(0)=\theta_2(0)$ and $\theta_0(0)-\theta_1(0)\in (-\pi,\pi+c(\alpha))\cup (\pi+c(\alpha),\pi)$, which means $x_1(0)=x_2(0)$ and $x_1(0)\in (-\pi,\pi+c(\alpha))\cup (\pi+c(\alpha),\pi)$. When $x_1(0)\in (-\pi,\pi+c(\alpha))$, we choose $V_1=\frac{1}{2}(x_1-c(\alpha))^2$ as a Lyapunov candidate. Its time derivative is  $\dot V_1=-A(x_1-c(\alpha))\big( 2\sin (x_i+\alpha) +\sin(x_i-\alpha)\big)$, which satisfies $\dot V<0$ for any $x_1\in (-\pi,\pi+c(\alpha))$ and $\dot V=0$ if $x_1=c(\alpha)$. Thus, starting from $(-\pi,\pi+c(\alpha))$, $x_1(t)$ converges to $c(\alpha)$ asymptotically. When $x_1(0)\in (\pi+c(\alpha),\pi)$, we choose $V_1=\frac{1}{2}(x_1-2\pi-c(\alpha))^2$ as a Lyapunov candidate. Likewise, one can show starting from $(\pi+c(\alpha),\pi)$, $x_1(t)$ converges to $2\pi+c(\alpha)$ asymptotically. Since $c(\alpha)$ and $2\pi+c(\alpha)$ represent the same point on $\RS^1$,  the two equilibrium points of \eqref{dynamics:M}, $x_1=x_2=c(\alpha)$ and $x_1=x_2=2\pi+c(\alpha)$, correspond to the same manifold $\mathcal M_1$ of $\theta\in\RS^3$. Then, there is no positively invariant manifolds in $\mathcal M$ other than $\mathcal M_1$ and $\mathcal M'_1$, since starting from any point  in $\mathcal M/\mathcal M'_1$, $\theta(t)$ converges to $\mathcal M_1$. 
	
	Second, it remains to study the stability of $\mathcal M_1$ and $\mathcal M_1'$ for different values of $\alpha$ since they are the only positively invariant manifolds in $\mathcal M$.
	The Jacobian matrix of  \eqref{linearized} evaluated at the $x=(x_1,x_2)^\top$ is
	\begin{align*}
	J(x)&={- \footnotesize{A}}  {\SmallMatrix{\cos \left( {{x_1} + \alpha } \right) + \cos \left( {{x_1} - \alpha } \right)&{\cos \left( {{x_2} + \alpha } \right)}\\{\cos \left( {{x_1} + \alpha } \right)}&\cos \left( {{x_2} + \alpha } \right) +  \cos \left( {{x_2} - \alpha } \right)}}.
	\end{align*}
	If $\alpha < \arctan (\sqrt{3} )$, all the eigenvalues of $J(c(\alpha))$ are negative, which proves that $\mathcal M_1$ is exponentially stable; on the other hand,  $J(\pi+c(\alpha))$ has a positive eigenvalue, which means $\mathcal M'_1$ is unstable. Then, there is a unique exponentially stable remote synchronization manifold, that is $\mathcal M_1$. Following similar lines, one can show  both $\mathcal M_1$ and $\mathcal M'_1$ are unstable if $\alpha > \arctan (\sqrt{3} )$, which proves 2).   This implies the remote synchronization manifold $\mathcal M$ is unstable if $\alpha > \arctan (\sqrt{3})$. 
\end{IEEEproof}

\subsection{Proposition \ref{dynamics:prop}}\label{proposition:04}
	\begin{proposition}\label{dynamics:prop}
	The dynamics of~$\mu(t)$ in~\eqref{distance} and~$\zeta (t)$ in~\eqref{expre:phi} are given by \eqref{expres:r:3} and \eqref{exre:dphi_dt:1}, respectively.
\end{proposition}
\begin{IEEEproof}
	We observe that $(\cos(\theta_0-\theta_1)+\cos(\theta_0-\theta_2))^2+(\sin(\theta_0-\theta_1)+\sin(\theta_0-\theta_2))^2=2+2\cos (\theta_0-\theta_1)\cos(\theta_0-\theta_2)+2\sin (\theta_0-\theta_1)\sin(\theta_0-\theta_2)$. Using the trigonometric identity $\cos \beta_1 \cos \beta_2+\sin \beta_1 \sin \beta_2= \cos(\beta_1-\beta_2)$ one obtains that $(\cos(\theta_0-\theta_1)+\cos(\theta_0-\theta_2))^2+(\sin(\theta_0-\theta_1)+\sin(\theta_0-\theta_2))^2=2+2\cos(\theta_1-\theta_2)$. Substituting this equality into \eqref{expre:r} we have
	\begin{align}\label{simplif:r}
	r=\frac{1}{2}\sqrt{2+2\cos(\theta_1-\theta_2)}= \sqrt{\cos^2\frac{\theta_1-\theta_2}{2}},
	\end{align}
	where the second equality has used the trigonometric identity $\cos 2\beta=2\cos^ 2 \beta-1$. The time derivative of $\mu(t)$ then satisfies
	\begin{align}
	&\frac{d\mu(t)}{dt}=-\frac{dr(t)}{dt} \nonumber\\
	&{ =\frac{1}{2\sqrt{\cos^2\frac{\theta_1-\theta_2}{2}}}2 \cos(\frac{\theta_1-\theta_2}{2}) \sin(\frac{\theta_1-\theta_2}{2}) \frac{\dot \theta_1 - \dot \theta_2}{2}}\nonumber\\
	&=\frac{1}{2r}\sin(\frac{\theta_1-\theta_2}{2}) \cos(\frac{\theta_1-\theta_2}{2}) (\dot \theta_1 - \dot \theta_2). \label{drdt:gneral:2}
	\end{align}
	It follows from \eqref{dyna_peripheral} that 
	\begin{align*}
	\dot \theta_1 - \dot \theta_2&=A\sin(\theta_0-\theta_1-\alpha)-A\sin(\theta_0-\theta_2-\alpha)\\
	&=-2A \sin( \frac{\theta_1-\theta_2}{2} )\cos( \theta_0-\frac{\theta_1+\theta_2}{2} -\alpha),
	\end{align*}
	where the second equality has used the trigonometric identity $\sin\beta_1-\sin \beta_2=2 \sin (\frac{\beta_1-\beta_2}{2})\cos (\frac{\beta_1+\beta_2}{2})$.
	Substituting this expression of $\dot \theta_1 - \dot \theta_2$ into \eqref{drdt:gneral:2} yields
	\begin{align}
	\frac{d \mu(t)}{d t}=&-\frac{A}{r} \sin^2 ( \frac{\theta_1-\theta_2}{2} )  \nonumber\\
	&\times \cos ( \frac{\theta_1-\theta_2}{2} )\cos (  \theta_0-  \frac{\theta_1+\theta_2}{2} -\alpha )  \nonumber\\
	=&-\frac{A}{2r} (1-r^2) \sum_{j=1}^{2} \cos(\theta_0-\theta_j-\alpha), \label{expres:r:2}
	\end{align}	
	where $\sin^2 ( \frac{\theta_1-\theta_2}{2})=1-r^2 $ and the trigonometric identity $\cos\beta_1\cos\beta_2=(\cos(\beta_1-\beta_2)+\cos(\beta_1+\beta_2))/2$ have been used to obtain the second equality.
	We further observe that
	\begin{align}
	&\sum_{j=1}^{2} \cos(\theta_0-\theta_j-\alpha) \nonumber\\
	&=\cos \alpha \sum_{j=1}^{2} \cos (\theta_0-\theta_j) +\sin \alpha \sum_{j=1}^{2} \sin (\theta_0-\theta_j).  \nonumber
	\end{align}
	From \eqref{nota:z}, there hold that $\sum_{j=1}^{2}\cos (\theta_0-\theta_j)=2 z_1$ and $\sum_{j=1}^{2}\sin (\theta_0-\theta_j)=2 z_2$. It then follows from \eqref{polar:cos} and \eqref{polar:sin} that
	\begin{align}
	&\sum_{j=1}^{2} \cos(\theta_0-\theta_j-\alpha) \nonumber\\
	&=2r \cos \alpha \cos  \zeta +2r \sin \alpha \sin  \zeta=2 r\cos( \zeta-\alpha).\label{cos:intermie}
	\end{align}	
	Substituting \eqref{cos:intermie} into \eqref{expres:r:2} one obtains
	\begin{align*}
	\frac{d \mu(t)}{d t}=-A(1-r^2)\cos(\zeta-\alpha),
	\end{align*}
	which is nothing but~\eqref{expres:r:3} since $r=1-\mu$.
	
	We next derive the time derivative of $ \zeta(t)$ given in \eqref{exre:dphi_dt:1}. It holds that $ \zeta=\arctan(z_2/z_1)$, and then the time derivative of $ \zeta$ satisfies
	\begin{align}
	\frac{d  \zeta(t)}{dt}=\frac{1}{z_1^2+z_2^2}(z_1 \dot z_2-z_2 \dot z_1). \label{exre:dphi_dt}
	\end{align}
	The time derivatives of $z_1$ and $z_2$ can be calculated using \eqref{nota:z}, and it then follows that
	\begin{align*}
	&z_1 \dot z_2-z_2 \dot z_1\\
	&=\frac{1}{4}\sum_{j=1}^{2}\cos (\theta_0-\theta_j) \Big(\sum_{j=1}^{2}\cos (\theta_0-\theta_j)\cdot(\dot \theta_0- \dot \theta_j)\Big)\\
	&+\frac{1}{4}\sum_{j=1}^{2}\sin (\theta_0-\theta_j)\Big(\sum_{j=1}^{2}\sin (\theta_0-\theta_j)\cdot(\dot \theta_0- \dot \theta_j)\Big)\\
	&=\frac{1}{4}\sum_{j=1}^{2}(\dot \theta_0- \dot \theta_j)+\frac{1}{4}\sum_{j=1}^{2}\Big(\cos (\theta_0-\theta_j)\cos (\theta_0-\theta_{- j})\\
	&+\sin (\theta_0-\theta_{j})\sin (\theta_0-\theta_{- j})\Big)\cdot(\dot \theta_0- \dot \theta_{- j}),
	\end{align*}
	where $-j$ is defined in a way so that $-j=2$ if $j=1$, and $-j=1$ otherwise. By using the trigonometric identity $\cos \beta_1\cos \beta_2+\sin \beta_1\sin \beta_2=\cos(\beta_1-\beta_2)$, we have 
	\begin{align*}
	z_1 \dot z_2-z_2 \dot z_1&=\frac{1}{4}\sum_{j=1}^{2}(\dot \theta_0- \dot \theta_j)+\frac{1}{4}\cos(\theta_1-\theta_2)\sum_{j=1}^{2}(\dot \theta_0- \dot \theta_j)\\
	&=\frac{1}{2}\cos^2\frac{\theta_1-\theta_2}{2}\sum_{j=1}^{2}(\dot \theta_0- \dot \theta_j)\\
	&=\frac{1}{2}r^2\sum_{j=1}^{2}(\dot \theta_0- \dot \theta_j),
	\end{align*}
	where the second equality has used the identity $1+\cos(\theta_1-\theta_2)=2\cos^2 (\frac{\theta_1-\theta_2}{2})$.
	It can be calculated by the system dynamics \eqref{dyna:3osci} that 
	\begin{align*}
	&\dot \theta_0-\frac{1}{2}(\dot \theta_1+\dot \theta_2)\\
	&=u+A\sum_{j=1}^{2}\sin (\theta_j-\theta_0-\alpha) -\frac{A}{2}\sum_{j=1}^{2}\sin(\theta_0-\theta_j-\alpha)\\
	&=u-A\cos \alpha \sum_{j=1}^{2}\sin (\theta_0-\theta_j)-A\sin \alpha \sum_{j=1}^{2}\cos (\theta_0-\theta_j)\\
	&\;\;\;\;\;\;\;-\frac{A}{2}\cos \alpha \sum_{j=1}^{2}\sin (\theta_0-\theta_j)+\frac{A}{2}\sin \alpha \sum_{j=1}^{2}\cos (\theta_0-\theta_j).
	\end{align*}
	Substituting \eqref{nota:z}  into the above equation, we obtain
	\begin{align*}
	&\dot \theta_0-\frac{1}{2}(\dot \theta_1+\dot \theta_2)=u-2A\cos \alpha z_2-2A\sin \alpha z_1\\
	&\;\;\;\;\;\;\;\;\;\;\;\;\;\;\;\;\;\;\;\;\;\;\;\;\;\;\;\;\;\;\;\;\;\;\;\;-A\cos\alpha z_2 +A\sin \alpha z_1\\
	&=u-2Ar(\cos\alpha \sin \zeta +\sin \alpha\cos \zeta)\\
	& \;\;\;\;\;\;\;\;\;\;\;\;\;\;\;\;\;\;\;\;\;\;\;\;\;\;\;\;\;\;\;-Ar(\cos\alpha\sin \zeta - \sin \alpha\cos\zeta)\\
	&=u-2Ar\sin( \zeta+\alpha)-Ar\sin(\zeta-\alpha),
	\end{align*}
	where the second equality has used the equations in \eqref{polar:cos} and \eqref{polar:sin}. As a consequence, $z_1 \dot z_2-z_2 \dot z_1=r^2(u-2Ar\sin(\zeta+\alpha)-Ar\sin(\zeta-\alpha))$. Using this inequality and the fact $z_1^2+z_2^2=r^2$ in \eqref{exre:dphi_dt}, we obtain 
	\begin{align*}
	\frac{d \zeta(t)}{dt}=u-Ar
	\big (2\sin(\zeta+\alpha)+\sin(\zeta-\alpha) \big),
	\end{align*}
	which is nothing but~\eqref{exre:dphi_dt:1} since $r=1-\mu$.
\end{IEEEproof}


%


\ifCLASSOPTIONcaptionsoff
  \newpage
\fi

\bibliographystyle{IEEEtran}
\bibliography{IEEEabrv,references}

\begin{thebibliography}{10}
\providecommand{\url}[1]{#1}
\csname url@samestyle\endcsname
\providecommand{\newblock}{\relax}
\providecommand{\bibinfo}[2]{#2}
\providecommand{\BIBentrySTDinterwordspacing}{\spaceskip=0pt\relax}
\providecommand{\BIBentryALTinterwordstretchfactor}{4}
\providecommand{\BIBentryALTinterwordspacing}{\spaceskip=\fontdimen2\font plus
\BIBentryALTinterwordstretchfactor\fontdimen3\font minus
  \fontdimen4\font\relax}
\providecommand{\BIBforeignlanguage}[2]{{%
\expandafter\ifx\csname l@#1\endcsname\relax
\typeout{** WARNING: IEEEtran.bst: No hyphenation pattern has been}%
\typeout{** loaded for the language `#1'. Using the pattern for}%
\typeout{** the default language instead.}%
\else
\language=\csname l@#1\endcsname
\fi
#2}}
\providecommand{\BIBdecl}{\relax}
\BIBdecl

\bibitem{vorotnikov1997partial}
V.~Vorotnikov, \emph{Partial {S}tability and {C}ontrol}.\hskip 1em plus 0.5em
  minus 0.4em\relax Springer Science \& Business Media, 1997.

\bibitem{haddad2011nonlinear}
W.~M. Haddad and V.~Chellaboina, \emph{Nonlinear {D}ynamical {S}ystems and
  {C}ontrol: {A} Lyapunov-{B}ased {A}pproach}.\hskip 1em plus 0.5em minus
  0.4em\relax Princeton, NJ, USA: Princeton University Press, 2011.

\bibitem{sinitsyn1991stability}
V.~Sinitsyn, ``Stability of the solution of a problem of inertial navigation,''
  \emph{Certain Problems of the Dynamics of Mechanical Systems}, pp. 46--50,
  1991.

\bibitem{willems1974partial}
J.~Willems, ``A partial stability approach to the problem of transient power
  system stability,'' \emph{International Journal of Control}, vol.~19, no.~1,
  pp. 1--14, 1974.

\bibitem{isidori1990output}
A.~Isidori and C.~I. Byrnes, ``Output regulation of nonlinear systems,''
  \emph{IEEE Transactions on Automatic Control}, vol.~35, no.~2, pp. 131--140,
  1990.

\bibitem{gambuzza2013analysis}
L.~V. Gambuzza, A.~Cardillo, A.~Fiasconaro, L.~Fortuna, J.~G{\'o}mez-Gardenes,
  and M.~Frasca, ``Analysis of remote synchronization in complex networks,''
  \emph{Chaos: An Interdisciplinary Journal of Nonlinear Science}, vol.~23,
  no.~4, p. 043103, 2013.

\bibitem{qin2018stability}
Y.~Qin, Y.~Kawano, and M.~Cao, ``Stability of remote synchronization in star
  networks of {K}uramoto oscillators,'' in \emph{57th IEEE Conference on
  Decision and Control}, 2018, pp. 5209--5214.

\bibitem{miroshnik2004attractors}
I.~V. Miroshnik, ``Attractors and partial stability of nonlinear dynamical
  systems,'' \emph{Automatica}, vol.~40, no.~3, pp. 473--480, 2004.

\bibitem{hancock2014restricted}
E.~J. Hancock and D.~J. Hill, ``Restricted partial stability and
  synchronization,'' \emph{IEEE Transactions on Circuits and Systems I: Regular
  Papers}, vol.~61, no.~11, pp. 3235--3244, 2014.

\bibitem{kuehn2015multiple}
C.~Kuehn, \emph{Multiple Time Scale Dynamics}.\hskip 1em plus 0.5em minus
  0.4em\relax Berlin: Springer, 2015.

\bibitem{al2009chaotic}
K.~Al-Naimee, F.~Marino, and et. al., ``Chaotic spiking and incomplete
  homoclinic scenarios in semiconductor lasers with optoelectronic feedback,''
  \emph{New Journal of Physics}, vol.~11, no.~7, p. 073022, 2009.

\bibitem{petrov1992mixed}
V.~Petrov, S.~K. Scott, and K.~Showalter, ``Mixed-mode oscillations in chemical
  systems,'' \emph{The Journal of Chemical Physics}, vol.~97, no.~9, pp.
  6191--6198, 1992.

\bibitem{chellaboina2002unification}
V.~Chellaboina and W.~M. Haddad, ``A unification between partial stability and
  stability theory for time-varying systems,'' \emph{IEEE Control Systems
  Magazine}, vol.~22, no.~6, pp. 66--75, 2002.

\bibitem{khalil2002nonlinear}
H.~K. Khalil, \emph{Nonlinear Systems,}.\hskip 1em plus 0.5em minus 0.4em\relax
  Upper Saddle River, NJ, USA: Prentice Hall, 2002.

\bibitem{aeyels1999exponential}
D.~Aeyels and J.~Peuteman, ``On exponential stability of nonlinear time-varying
  differential equations,'' \emph{Automatica}, vol.~35, no.~6, pp. 1091--1100,
  1999.

\bibitem{tsinias2013averaging}
J.~Tsinias and A.~Stamati, ``Averaging criteria for asymptotic stability of
  time-varying systems,'' \emph{IEEE Transactions on Automatic Control},
  vol.~59, no.~6, pp. 1654--1659, 2013.

\bibitem{varela2001brainweb}
F.~Varela, J.-P. Lachaux, E.~Rodriguez, and J.~Martinerie, ``The brainweb:
  phase synchronization and large-scale integration,'' \emph{Nature Reviews
  Neuroscience}, vol.~2, no.~4, p. 229, 2001.

\bibitem{qin2019partial}
Y.~Qin, Y.~Kawano, O.~Portoles, and M.~Cao, ``Partial phase cohesiveness in
  networks of {K}uramoto oscillator networks,'' \emph{arXiv preprint
  arXiv:1906.01065}, 2019.

\bibitem{menara2019stability}
T.~Menara, G.~Baggio, D.~Bassett, and F.~Pasqualetti, ``Stability conditions
  for cluster synchronization in networks of heterogeneous {K}uramoto
  oscillators,'' \emph{IEEE Transactions on Control of Network Systems}, 2019.

\bibitem{dorfler2014synchronization}
F.~D{\"o}rfler and F.~Bullo, ``Synchronization in complex networks of phase
  oscillators: A survey,'' \emph{Automatica}, vol.~50, no.~6, pp. 1539--1564,
  2014.

\bibitem{nicosia2013remote}
V.~Nicosia, M.~Valencia, M.~Chavez, A.~D{\'\i}az-Guilera, and V.~Latora,
  ``Remote synchronization reveals network symmetries and functional modules,''
  \emph{Physical Review Letters}, vol. 110, no.~17, p. 174102, 2013.

\bibitem{vicente2008dynamical}
R.~Vicente, L.~L. Gollo, C.~R. Mirasso, I.~Fischer, and G.~Pipa, ``Dynamical
  relaying can yield zero time lag neuronal synchrony despite long conduction
  delays,'' \emph{Proceedings of the National Academy of Sciences}, vol. 105,
  no.~44, pp. 17\,157--17\,162, 2008.

\bibitem{fischer2006zero}
I.~Fischer, R.~Vicente, J.~M. Buld{\'u}, M.~Peil, C.~R. Mirasso, M.~Torrent,
  and J.~Garc{\'\i}a-Ojalvo, ``Zero-lag long-range synchronization via
  dynamical relaying,'' \emph{Physical Review Letters}, vol.~97, no.~12, p.
  123902, 2006.

\bibitem{wagemakers2007isochronous}
A.~Wagemakers, J.~M. Buld{\'u}, and M.~A. Sanju{\'a}n, ``Isochronous
  synchronization in mutually coupled chaotic circuits,'' \emph{Chaos: An
  Interdisciplinary Journal of Nonlinear Science}, vol.~17, no.~2, p. 023128,
  2007.

\bibitem{banerjee2012enhancing}
R.~Banerjee, D.~Ghosh, E.~Padmanaban, R.~Ramaswamy, L.~Pecora, and S.~K. Dana,
  ``Enhancing synchrony in chaotic oscillators by dynamic relaying,''
  \emph{Physical Review E}, vol.~85, no.~2, p. 027201, 2012.

\bibitem{gambuzza2016inhomogeneity}
L.~V. Gambuzza, M.~Frasca, L.~Fortuna, and S.~Boccaletti, ``Inhomogeneity
  induces relay synchronization in complex networks,'' \emph{Physical Review
  E}, vol.~93, no.~4, p. 042203, 2016.

\bibitem{sakaguchi1986soluble}
H.~Sakaguchi and Y.~Kuramoto, ``A soluble active rotater model showing phase
  transitions via mutual entertainment,'' \emph{Progress of Theoretical
  Physics}, vol.~76, no.~3, pp. 576--581, 1986.

\bibitem{panaggio2015chimera}
M.~J. Panaggio and D.~M. Abrams, ``Chimera states: Coexistence of coherence and
  incoherence in networks of coupled oscillators,'' \emph{Nonlinearity},
  vol.~28, no.~3, p. R67, 2015.

\bibitem{hoppensteadt2012weakly}
F.~C. Hoppensteadt and E.~M. Izhikevich, \emph{Weakly Connected Neural
  Networks}.\hskip 1em plus 0.5em minus 0.4em\relax New York, USA: Springer
  Science \& Business Media, 2012.

\bibitem{adams2009rotating}
M.~L. Adams, \emph{Rotating Machinery Vibration: From Analysis to
  Troubleshooting}.\hskip 1em plus 0.5em minus 0.4em\relax New York, NY, USA:
  CRC Press, 2009.

\bibitem{wertz2012spacecraft}
J.~R. Wertz, \emph{Spacecraft Attitude Determination and Control}.\hskip 1em
  plus 0.5em minus 0.4em\relax Amsterdam, the Netherlands: Reidel, 1978.

\bibitem{terrell2009stability}
W.~J. Terrell, \emph{Stability and stabilization: An introduction}.\hskip 1em
  plus 0.5em minus 0.4em\relax New Jersey, USA: Princeton University Press,
  2009.

\end{thebibliography}

\end{document}